# The Interaction Between Domestic Monetary Policy and Macroprudential Policy in Israel[*]

Jonathan Benchimol,[‡] Inon Gamrasni,[‡] Michael Kahn,[‡] Sigal Ribon,[‡] Yossi Saadon,[‡] Noam Ben-Ze'ev,[‡] Asaf Segal,[‡] and Yitzchak Shizgal[‡]

July 2022

**Abstract**

The global financial crisis (GFC) triggered the use of macroprudential policies imposed on the banking sector. Using bank-level panel data for Israel for the period 2004–2019, we find that domestic macroprudential measures changed the composition of bank credit growth but did not affect the total credit growth rate. Specifically, we show that macroprudential measures targeted at the housing sector moderated housing credit growth but tended to increase business credit growth. We also find that accommodative monetary policy surprises tended to increase bank credit growth before the GFC. We show that accommodative monetary policy surprises increased consumer credit when interacting with macroprudential policies targeting the housing market. Accommodative monetary policy interacted with nonhousing macroprudential measures to increase total credit.

[*] Any views expressed in the paper are those of the authors and do not necessarily reflect those of the Bank of Israel. The authors are grateful to Sushanta Mallick (Editor), the associate editor, anonymous referees, Steven Laufer, and participants at the Bank of Israel Research Department seminar and the International Banking Research Network (IBRN) Meeting, Federal Reserve Bank of New York (2019), for their helpful comments. This research was initiated as part of the 2018 IBRN project examining the interaction between macroprudential policy and monetary policy.
[‡] Bank of Israel, Jerusalem, Israel. Corresponding authors: inon.gamrasni@boi.org.il; sigal.ribon@boi.org.il.

# 1 Introduction

House prices in Israel have more than doubled since 2008, alongside a substantial increase in housing credit (mortgages). This development, which may pose a significant risk to the financial sector and particularly to banks, together with an understanding of the importance of the financial sector stability as a whole, triggered the imposition of macroprudential (MaP) measures on the banks. This paper analyzes the effects of MaP policy and domestic monetary policy on domestic bank credit in Israel between 2004 and 2019. We chose to investigate the effects of MaP policy on housing, consumer (nonhousing), and business sector credit separately in order to understand the direct effect of the MaP measures on the market at which they were targeted and possible effects on the other bank credit sectors and on total credit.

We find that alongside the desired direct, moderating effect of the MaP measures on the credit markets at which they were targeted, the effect of these measures on total bank credit was weakened due to substitution effects. We show that the MaP measures targeted at housing credit reduced its growth rate, but tended to increase the growth rate of business credit. MaP measures targeted at reducing banks' credit concentration did not significantly affect consumer or business credit, but contributed to the growth of housing credit. We find that other general MaP measures (capital requirements from the banks) that were expected to slow the growth of business credit influenced business credit with a significant lag, apparently due to the time frame given to the banks to complete the adjustment of capital and risk assets. While the interaction between accommodative monetary policy and MaP measures targeting the housing market did not have any additional effect on housing credit, we find weak support for its expansionary effect on consumer credit. Overall, MaP measures did not change the development of total credit significantly, but they did change its composition. Monetary policy was generally found to be effective, with less accommodative policy slowing housing and consumer credit growth only before the global financial crisis (GFC). In addition, we find that higher foreign policy rates tended to reduce housing and consumer credit growth, similar to the impact of domestic policy, suggesting their ability to indicate future domestic monetary policy.

MaP policy is a regulatory tool that seeks to ensure the financial system's stability, which is necessary for stable economic growth. The overall objective of MaP policy is to limit the buildup of financial vulnerabilities in order to reduce the financial system's sensitivity to shocks. MaP policy targets a specific credit sector, such as the consumer, business, or housing credit sector, to achieve a particular objective. Specifically, MaP measures targeting housing credit are intended to limit banks' exposure to mortgage defaults. Consequently, most of these measures are expected to increase the mortgage cost, moderating housing credit growth. A similar mechanism applies to capital adequacy or banking concentration MaP measures. While MaP measures tend to decrease the growth rate of the targeted credit, substitution effects may be reflected in the growth of other credit sectors. For instance, MaP measures targeting housing credit may increase consumer credit as the latter may serve as an alternative to the former, whether for banks as suppliers of credit or for borrowers as demanders of credit. Alternatively, consumer credit may decline if MaP measures decrease the demand for housing, thus reducing the demand for housing-related consumption and consumer credit. Another example relates to MaP measures that limit credit to large companies. The supply of this credit may move to smaller businesses but can also be routed to housing or consumer credit.

The use of MaP measures has expanded since the GFC (Cerutti et al., 2017). The theoretical literature based on dynamic stochastic general equilibrium (DSGE) models supports the view that MaP measures can dampen credit cycles. Using a DSGE model with a housing sector, Kannan et al. (2012) demonstrated that monetary and macroeconomic policies can help stabilize the economy, given the shocks that hit the economy. When financial or housing demand shocks drive a credit and housing boom, MaP measures can temper them and improve welfare. Nevertheless, DSGE models offer limited empirical guidance on the influence of the interaction between MaP measures and monetary policy on credit cycles. We aim to assess these empirical linkages. Repullo and Suarez (2013) showed that the credit (financial) and business cycles are not necessarily synchronized. Therefore, monetary policy alone cannot smooth both cycles. Consequently, MaP policy plays a more significant role in moderating the financial cycle according to their analysis, thus, promoting financial stability, while monetary policy remains responsible for price and output stability.

The effects of the interaction between MaP and monetary policies is still unsettled (Dell'Ariccia et al., 2012; IMF, 2013; Aiyar et al., 2014). De Marco et al. (2021) found that stricter monetary policy and MaP measures reinforced each other for the UK, but only for small banks. Similarly, Forbes et al. (2017) found that monetary policy can amplify the impact of regulatory measures. By contrast, De Jonghe et al. (2020) found a trade-off between monetary policy and supervisory capital requirements in the case of Belgium: the higher the banks' capital requirements are, the weaker the effect of monetary accommodation on credit supply is. Cerutti et al. (2017) found that MaP measures have less of an effect on credit growth in advanced economies that tend to have alternative sources of nonbank credit and enable borrowers to obtain funds from abroad in more open economies. We seek to understand how MaP policy affects each component of bank credit, and whether the interaction with monetary policy has an additional effect on these credit sectors.

Gambacorta et al. (2020) examined the effect of domestic MaP measures on domestic credit through an empirical exercise similar to ours. Unlike in our estimation, they used granular information on bank loans at the firm level from eight different countries. They found that MaP measures are successful in easing credit cycles and reducing banking sector risk. They also found that bank-specific characteristics influence the impact of MaP measures on credit. Finally, they found that MaP measures that reinforce monetary policy (i.e., in the same direction of influence) are more effective.

Everett et al. (2021) contributed to the literature with Dutch and Irish data.[1] They employed a similar methodology to the one we adopt, examining the effect of monetary policy and MaP policy and their interaction on domestic mortgage lending. As both Ireland and the Netherlands are small economies in the euro system, monetary policy shocks may be regarded as exogenous to these two countries. Like Israel, the banking sector in Ireland and the Netherlands is relatively concentrated and both have employed various MaP measures in the past decade against the background of rapid increases in house prices. Everett et al. (2021) complemented their analysis by using confidential bank-level data for domestic lending. They found that positive (less accommodative) domestic monetary policy shocks reduced mortgage lending growth in both countries, while prudential regulations weakened this effect in Ireland but not in the Netherlands. They found only weak evidence for an international lending channel: a foreign (US, UK) monetary policy shock, with or without interaction with the MaP measures, was found to have no significant effect on mortgage lending for Ireland, while for the Netherlands, weak evidence was found that tightening monetary policy in the UK reduces mortgage lending.

---

[1] Everett et al. (2021), Cao et al. (2021), and Bussière et al. (2021) were all part of the 2018 International Banking Research Network (IBRN) project.

Cao et al. (2021) studied the interaction between MaP policy and monetary policy in Norway and Sweden and found evidence that aggregate MaP policy depresses aggregate lending, although different types of MaP measures do not lead to significantly different effects. Bussière et al. (2021) summarized the main findings of seven papers that related mainly to MaP policies targeted at capital flows and the possible spillovers from monetary policies in core economies to recipient economies. The findings show that MaP policies in recipient countries can partly offset monetary policy spillover effects from the core (large) economies. Bussière et al. (2021) also remarked that the impact differs considerably across MaP policy instruments, suggesting the importance of a detailed analysis.

In the absence of any interest rate adjustment, a credit-restricting MaP shock could reduce the debt to GDP ratio in the short term, but it is unlikely to reduce the ratio of firms' debt to their internal funds. Greenwood-Nimmo and Tarassow (2016) suggested that MaP policy alone is likely to have ambiguous effects on financial stability. Nevertheless, MaP policy is rarely used alone. When monetary policy adjusts in response to an MaP shock, both ratios decrease substantially, indicating an increase in financial stability. According to this finding, a combination of monetary and MaP measures for achieving financial stability may be desirable, as Yellen (2014) advocated. Arena et al. (2020) highlighted the possibility that MaP policies can limit riskier mortgages and contribute to stronger bank balance sheets, countering the accommodative monetary policy stance. Following this line of thought, we analyze the interaction between MaP and monetary policies and their effect on credit growth.

Following Altavilla et al. (2020), we show that monetary policy and MaP measures are mutually reinforcing in relation to consumer loans more than to mortgage loans. This result may be related to the risk-weighted ratio of capital to assets. Part of the banks' income can be used to rebuild capital and increase headroom if the latter is below its normal level, and margins on consumer loans can be raised to speed up the process (Davis et al., 2019).

As a small open economy, more than half of Israel's trades are related to the US and the Eurozone. Imported inflation and capital flows from these places directly influence domestic monetary policy and thus credit aggregates. Our findings that higher foreign policy rates moderate housing and consumer credit growth while tending to increase business sector credit growth complement the findings of Niepmann et al. (2020). Although foreign credit is not significant in Israel, our results indicate the importance of foreign monetary policy surprises for local credit aggregates.

The remainder of the paper is organized as follows. Section 2 briefly presents the economic background and the Israeli environment. Section 3 describes the data used in Section 4 for the estimations and results. Section 5 presents robustness checks and Section 6 concludes. The Appendix presents additional results.

## 2 Economic and Policy Background

### 2.1 Macroeconomic Background

Since the beginning of the new millennium, particularly after 2003 (and until 2019), the Israeli economy has enjoyed a relatively stable macroeconomic environment characterized by high growth rates and price stability. The economy was hit during these years by two major crises. The first, at the end of 2001, was due to the dot-com crisis coupled with local geopolitical hostilities. The second was the GFC, which the Israeli economy weathered quite well, experiencing only a short period of slowdown at the end of 2008 and the beginning of 2009 before returning to growth rates close to its potential in the second half of 2009 (Figure 1).

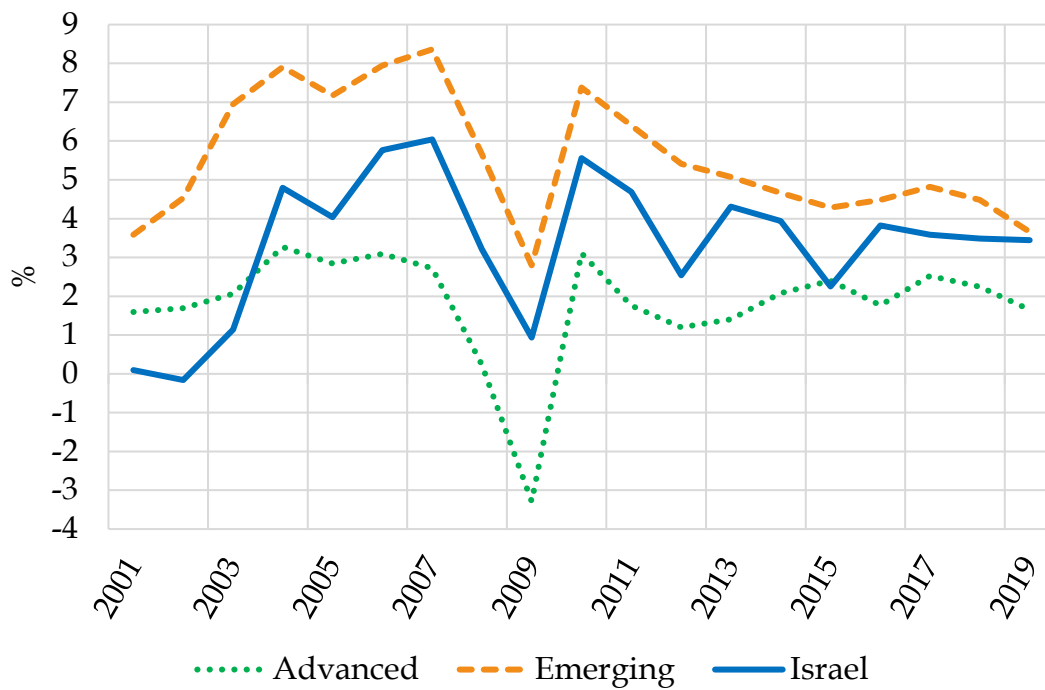

**Figure 1: Annual GDP Growth in Israel and Advanced and Emerging Economies**

*Notes*: Annual GDP growth in percent. *Sources*: Bloomberg and Central Bureau of Statistics.

Compared with the crisis' adverse effect on other advanced economies overall, the impact in Israel was relatively mild, more akin to the experience of developing economies. In particular, the crisis had a limited and moderate effect on Israel's financial system, and the financial institutions remained stable. The factors contributing to the GFC's relatively mild effect on Israel include a conservative financial system (and in particular a conservative and closely supervised banking system), a relatively balanced housing market, and a successful economic policy (both fiscal and monetary). Since the beginning of the 2000s, Israel has enjoyed low inflation rates within the framework of an inflation-targeting regime, with the target set between 1 and 3 percent, and actual inflation fluctuated around this target range. In recent years, inflation has been lower than the target, and monetary policy is generally characterized as accommodative (see details in Section 2.4).

## 2.2 Housing Sector

Israel enjoyed a vast influx of immigrants during the 1990s, increasing its population by about 20 percent (1 million people) within five years. The housing market—the private sector, accompanied by government intervention—reacted, and housing starts surged at the beginning of the 1990s, with prices increasing at double-digit rates. The cycle ended in the early 2000s, with building activity and house prices declining by about 12 percent from peak to trough. Between 2007 and 2018, house prices rose steadily, with the price level more than doubling within a decade (Figure 2).

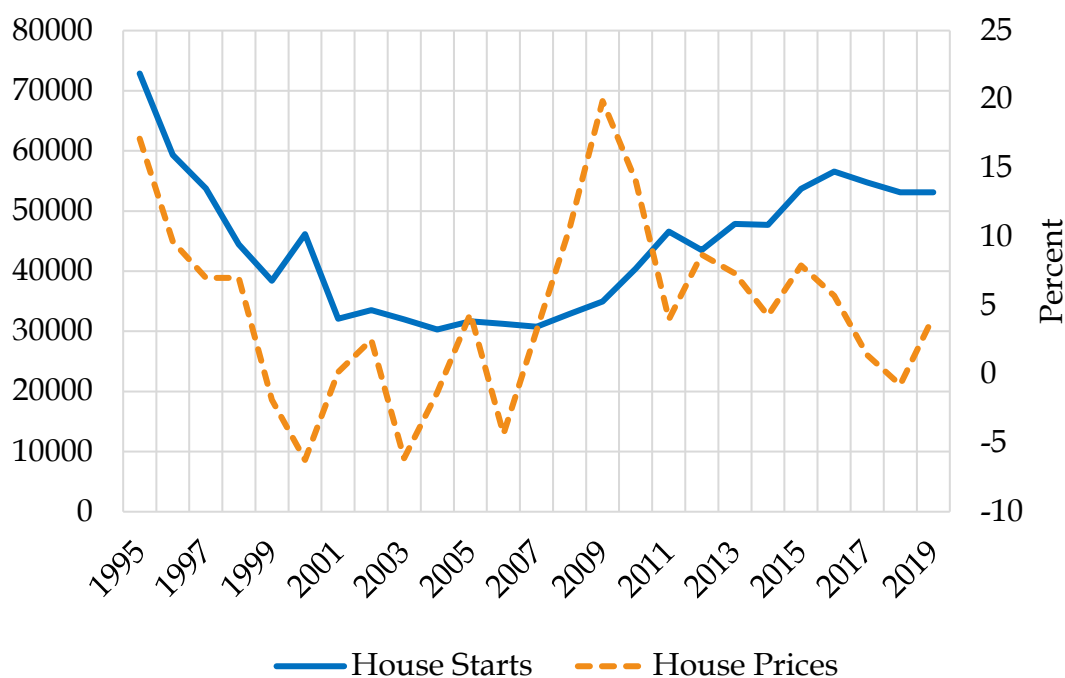

Figure 2: House Starts and House Prices in Israel

*Notes*: The left axis relates to the housing starts in units, and the right axis relates to nominal house prices, which are expressed in annual percentage changes. *Source*: Central Bureau of Statistics.

On the supply side, building starts reacted sluggishly to the increase in demand arising from both demographic needs and the rise in income and wealth, contributing to the rapid price increases.[2]

The housing supply side in Israel is generally inelastic. The planning process in Israel is very long and takes about 13 years, on average. This slow supply response results in large price effects in response to positive demand shocks.

A significant government program that was launched in 2015, which essentially subsidizes new apartments for buyers who meet specific characteristics (young first-time buyers), has had a significant impact on the market. This is because it removes both supply and demand from the free market, while in contrast, it may encourage households that did not plan to purchase a home to enter the market. Even before that, taxation had been changed to reduce investor demand.

---
[2] See the Bank of Israel Annual Reports for 2017 and 2018.

In the last few years of our sample, signs of a market slowdown have emerged, with declines in both housing starts and transactions and prices moderating and even declining for several quarters. The annual price increase in 2019 was about 3.5 percent (Figure 2).

As noted above, since 2008, house prices in Israel have increased nominally by about 130 percent. This rate of increase was exceptional compared to other countries (Figure 3). Looking at a different starting point partially reduces Israel's deviations from other countries. Comparing recent price-to-income ratios with past ones strongly warns of relatively high house prices in Israel as of the end of 2019.

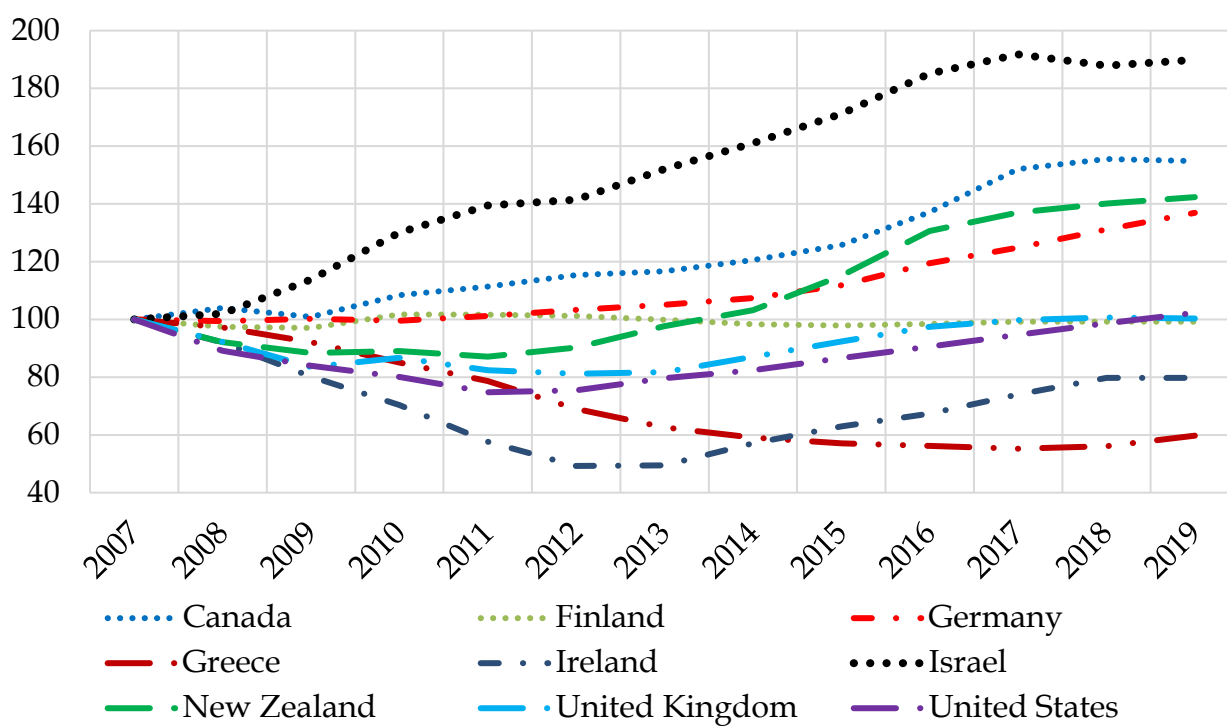

**Figure 3: House Prices in Israel and Various Countries**

*Notes*: The house price index is 100 in 2007. *Source*: OECD

## 2.3 Banking Sector

The Israeli banking sector is relatively concentrated and consists of a small number of banks. Seven banking groups control most of the domestic banking system, representing 99.6 percent of the banking sector, with two dominant banking groups (Bank Leumi and Bank Hapoalim) accounting for more than 50 percent of the market.

In our analysis, we focus on the credit portfolio of these seven banks.[3] The concentrated structure of the Israeli banking sector may be attributed to the small size of the Israeli economy, together with the existence of economies of scale and scope.

---

[3] Leumi, Hapoalim, Discount, Mizrahi-Tefahot, First International, Union Bank, and Bank of Jerusalem.

Due to conservative management and very close supervision, the banks in Israel exhibited substantial resilience to the developments of the GFC in the global financial system.

In the past, not only the banking system suffered from a high level of concentration, but the entire credit market did as well. Yet, the past two decades have seen changes due to the government reducing its primary borrower role in the economy. This change shifted private capital from investments in government bonds to private sector investments, thus easing big companies' access to credit from the capital markets, creating a movement toward nonbank credit. In addition, the number of loans given by institutional investors has grown, thanks to changes in the array of pension plans in Israel. These loans also contribute to the reduction of the banks' share of business and consumer credit (Figure 4).

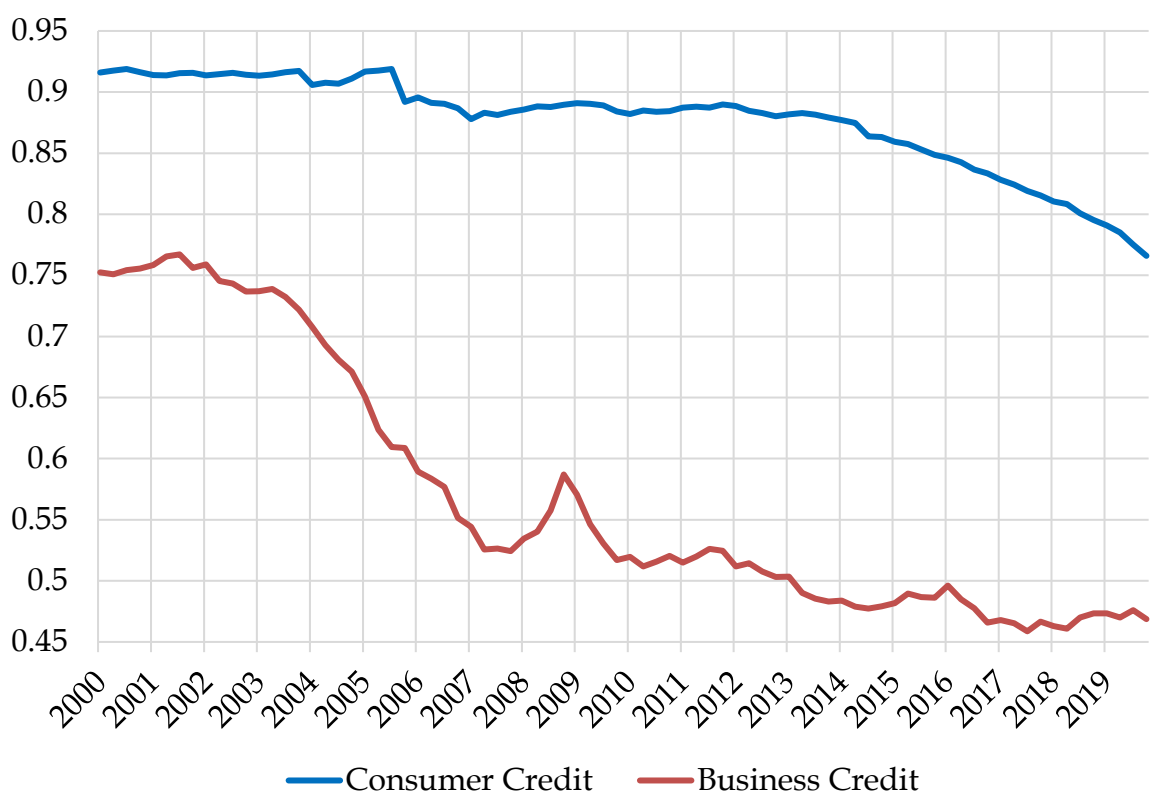

**Figure 4: Banks' Share of Business and Consumer Credit**

*Notes*: Percentage of consumer and business credit issued by private banks in Israel. *Source*: Bank of Israel, Banking Supervision Department.

The domestic credit portfolio among the seven big banks, which includes business (nonfinancial) credit and credit to households, has been growing since 2001. The largest component is the household sector, which includes rising levels of housing loans. In the fourth quarter of 2019, the total amount of credit stood at NIS 970 billion, the equivalent of approximately 69 percent of Israel's annual GDP (Figure 5a).

The share of housing credit in GDP has grown in view of increasing house prices. Consumer credit has remained relatively stable relative to GDP, and credit to large businesses has declined (Figure 5b).

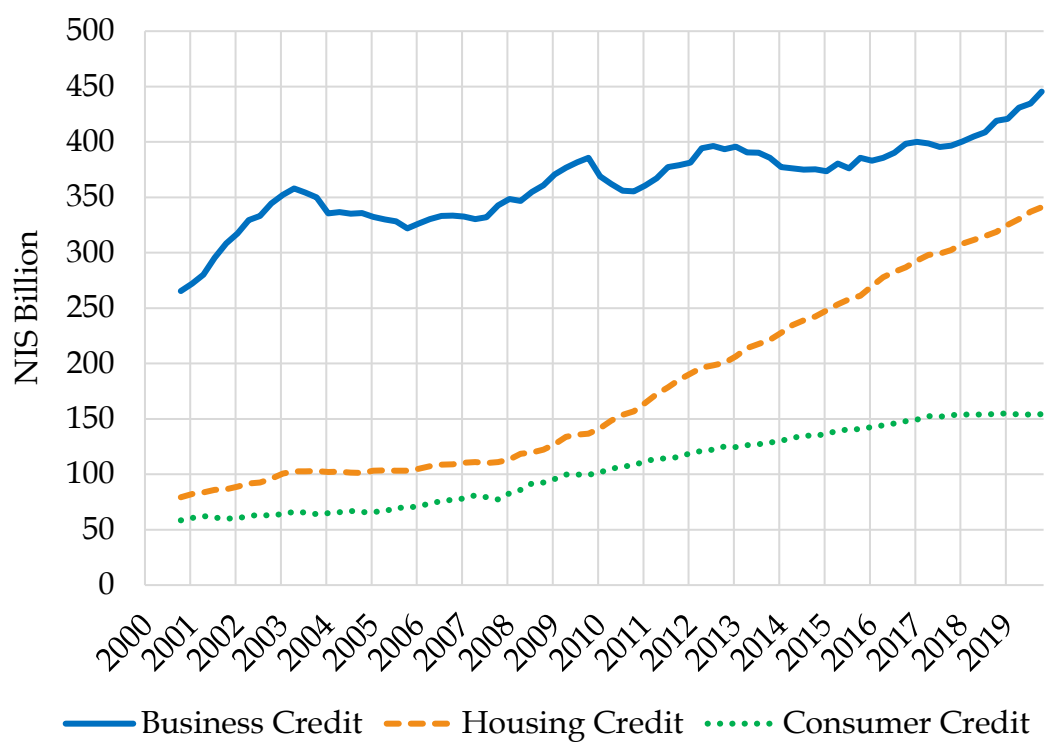

*Notes*: Credit balance of the seven major banks in Israel in NIS billion. *Source*: Bank of Israel, Banking Supervision Department.

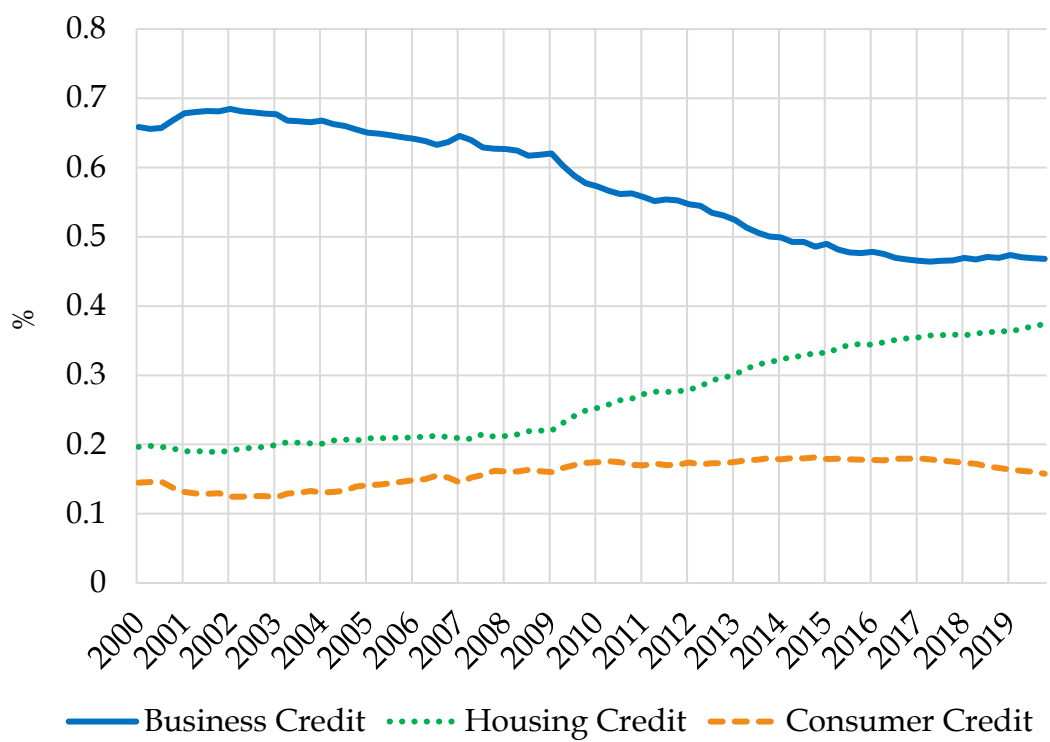

*Notes*: Ratio of business, housing, and consumer credit to real GDP of the seven major banks in Israel. *Source*: Bank of Israel, Banking Supervision Department.

On the side of business credit, the reduction in credit to GDP ratio can be linked to factors such as (1) requiring banks to reach regulatory capital targets by decreasing big business credit's share of total credit, (2) setting limits that should minimize concentration among borrowers, and (3) increasing access to institutional and market lending.

## 2.4 Monetary Policy

Since 2003, monetary policy in Israel has been conducted within the framework of a "flexible inflation target" regime, with the target set within a 1–3 percent band, and allowing for temporary deviations from the target range. The exchange rate regime may be characterized as a "managed float," as the Bank of Israel has been intervening in the foreign exchange market since March 2008—at first with pre-announced daily fixed amounts and since August 2009 in a discretionary manner, according to market conditions (Caspi et al., 2022).[4]

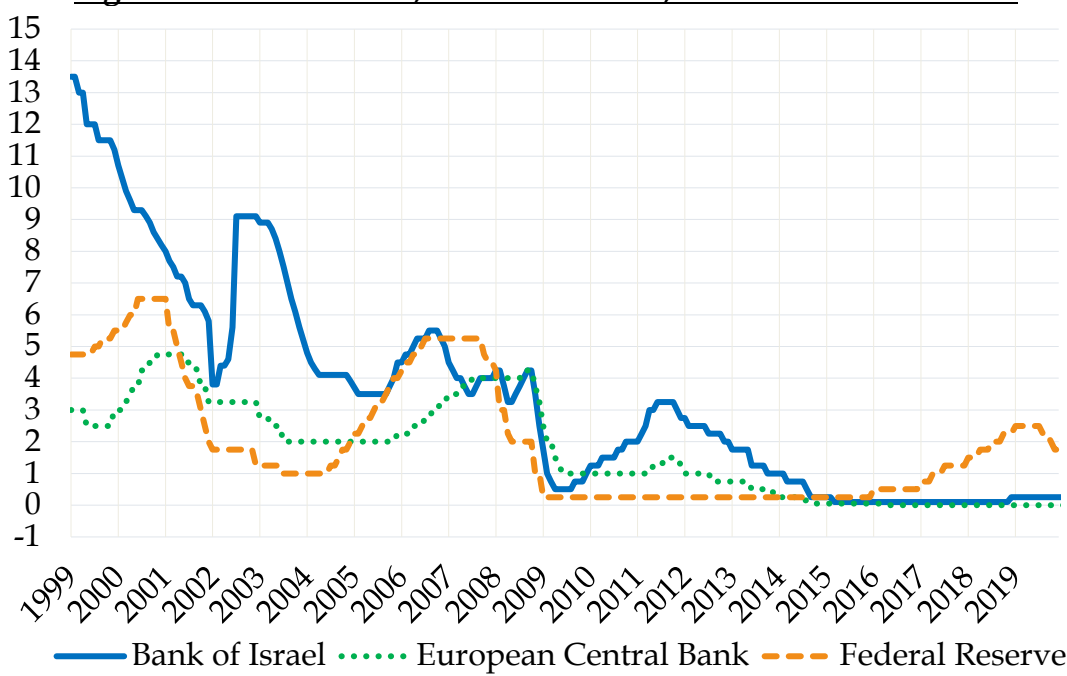

**Figure 6: Bank of Israel, Federal Reserve, and ECB Interest Rates**

*Notes*: Nominal interest rates. *Sources*: Bank of Israel, European Central Bank, and Board of Governors of the Federal Reserve System.

Since the GFC, domestic monetary policy has been very accommodative. The Bank of Israel did not set negative interest rates but, as noted above, it did use intervention in the foreign exchange market and government bond purchases for a short period as additional tools, and forward guidance to communicate the future monetary policy path better. The interest rate remained the principal monetary policy instrument. It was set to 0.1 percent in 2015, and increased to 0.25 percent in November 2018 (Figure 6).

---

[4] In addition, the Bank of Israel purchased a pre-announced amount annually as part of a program intended to offset the effects of natural gas production on the exchange rate. This program ceased at the beginning of 2019.

Looking farther into the past, the decline in the Bank of Israel's nominal interest rate reflected both the adaptation to a lower inflation environment and a decline in real interest rates, from about 6 percent at the beginning of the 2000s to about 2 percent a decade into the millennium, and to around zero in recent years, in line with the very accommodative global monetary policy.

## 2.5 Regulation and Macroprudential Measures

The debt to GDP ratio in Israel is low relative to other countries. Total debt to GDP is about 180 percent, with household debt (housing and consumer) to GDP particularly low at around 40 percent in 2019, with no significant change since 2000. Total business sector debt (excluding financial institutions) to GDP was about 70 percent in 2019, with bank credit constituting about half of this credit. Public debt to GDP in 2019 was around 60 percent.

In the past decade, the risk stemming from the housing market was the main concern for Israel's financial stability. Of particular note was the banks' large exposure to the housing market, comprising housing credit that has grown considerably over the past decade (Figure 5a), and credit exposure to the construction and real estate sectors, which together account for 47 percent of total bank credit. Against the background of the low interest rates since the GFC, households took a large portion of housing loans in a floating interest rate, which was cheaper than a fixed interest rate. However, the floating interest rate exposed households to risks stemming from a possible rise in market interest rates.

The Bank of Israel began to implement MaP measures at a relatively early stage of the financial cycle, about two years after the house price cycle started in 2008. Although the MaP measures targeting the housing sector were substantial, other MaP classes were also used in Israel during the past two decades. Some of them aimed to reduce the risks arising from the concentration of the banks' credit portfolio. Others aimed to reduce general risks of the banks and were associated with the Basel regulations. We divide these MaP measures into three major groups.[5] In the following sections, we refer to the measures taken in each group. A summary of all MaP measures taken by the Bank of Israel and included in these groups is presented in Table 1.

### 2.5.1 Housing Market Measures

MaP measures in the housing sector contain LTV-oriented measures aimed at influencing the quality of mortgages and borrowers' risk, and thus the level of risk to the banks from the mortgage market, and other measures aimed at strengthening the banks' resilience to a crisis, such as increased provisions and additional capital requirements (Claessens et al., 2013; Aiyar et al., 2014).

---

[5] See Section 3.4 for an explanation of the classification.

The first measures that were implemented in 2010 were capital requirements for risky loans, and provisions and restrictions on the characteristics of the loans that can be offered by the banks were set.[6]

In addition, a requirement for a supplementary capital allowance regarding housing loans was introduced. The risk-weighted assets (RWA) capital allocation criterion is 100 percent for loans of groups of borrowers who buy properties collectively. Later, in two consecutive measures (in July and October 2010), the RWA was set to 100 percent for loans with an LTV greater than 60 percent when the variable-rate credit component comprises more than 25 percent and the level of the loan is over NIS 800,000.

Regarding the MaP measure taken in October 2010—restrictions on the level of LTV—on house prices and borrowers, Baudot-Trajtenberg et al. (2017) found that the impact on house prices was limited, since only 15 percent of mortgage borrowers were affected by this move. The measure incentivized riskier borrowers to reduce leverage with an LTV of over 60 percent. In response to these restrictions, borrowers bought less expensive homes, farther from the center of the country, and in poorer neighborhoods.

Between 2011 and 2014, several MaP measures were adopted to strengthen banks' and households' stability. Although influencing house prices was not a declared goal of the Bank of Israel, moderated house prices would have been a welcome outcome.

In 2011, a limitation stating that only up to one-third of the total loan can bear interest at a variable rate that is adjusted at a frequency of up to 5 years was established. This measure was introduced to cope with the high demand for variable-rate loans in view of low interest rates, which led to new loans with variable interest rates reaching 80 percent of total new loans at that time.

Due to the continuing rise in housing prices and mortgage volumes, additional measures were implemented to limit exposure to mortgage loans. In October 2012, the LTV ratio was limited to 70 percent, except for first-time homebuyers (75 percent) and investors, including nonresidents (50 percent).

In March 2013, risk weights on some housing loans increased, and at the end of August 2013, four new measures were introduced: the PTI (payment-to-income) ratio was limited to 50 percent of income; housing loans where the monthly repayment is over 40 percent of income were weighted at 100 percent of RWA to calculate the capital adequacy ratio; the portion of the loan at a variable rate was limited to two-thirds of the total loan, and the portion of the loan at a variable rate that can change within five years was limited to one-third of the total loan; the loan period was limited to 30 years (inclusive).

In September 2014, an additional capital allowance requirement was introduced, amounting to 1 percent of outstanding housing loans, to address vulnerabilities and boost the banks' loss absorption capacity. Until 2017, house prices continued to rise despite the measures taken. The housing market remained very active, mortgage interest rates continued to decline, and the volume of mortgage lending rose, as did the volume of housing market transactions.

---

[6] In 2009 banks were warned to inform their customers of risks related to loans with variable interest rates. This warning is not included in the list of formal MaP measures that were implemented.

# Table 1: Macroprudential Measures

|   | Stability Measures | Decision Date | Application Date | Influence Date | Type |
|---|---|---|---|---|---|
| 1 | The directive regarding limitations on the indebtedness of a borrower and a group of borrowers was amended. | 20/08/2003 | 31/03/2004 | 20/08/2003 | CON |
| 2 | A 100% capital surcharge on groups of borrowers who buy properties collectively. | 25/03/2010 | 01/04/2010 | 25/03/2010 | HSE |
| 3 | Banking corporations shall make a supplemental provision at the rate of at least 0.75 percent on account of outstanding housing loans that were issued on or after July 1, 2010, and in which the existing ratio in each case between the debt (prorated to the bank's share of the mortgage) and the value of the mortgaged property on the date of loan execution exceeds 60 percent. | 11/07/2010 | 01/07/2010 | 11/07/2010 | HSE |
| 4 | Risk weights increased on housing loans with LTVs above 60%, a floating component of more than 25%, and a mortgage value greater than NIS 800,000. A minimum risk weight of 100% is applied to these mortgages (banks can lower it to 75%). This is a change from the previous 35–75% range. | 28/10/2010 | 26/10/2010 | 26/10/2010 | HSE |
| 5 | The core capital ratio target shall be set to a rate no lower than 7.5 percent. | 30/06/2010 | 31/12/2010 | 30/06/2010 | GEN |
| 6 | Limiting variable interest to a maximum of one-third of the mortgage loan. | 03/05/2011 | 05/05/2011 | 05/05/2011 | HSE |
| 7 | Limiting LTV ratio in housing loans: up to 75% for first-time homebuyers, up to 50% for investors, and up to 70% for those upgrading their homes. | 01/11/2012 | 01/11/2012 | 01/11/2012 | HSE |
| 8 | Risk weights on some housing loans are increased, depending on their LTV. Loans with LTVs between 45% and 60% have a higher risk weight of 50%, and those with LTVs between 60% and 75% are weighted at 75%. | 21/03/2013 | 01/01/2013 | 01/01/2013 | HSE |
| 9 | A bank will not issue a housing loan with a PTI of more than 50%. | 29/08/2013 | 01/09/2013 | 01/09/2013 | HSE |
| 10 | Risk weight of 100% imposed on mortgages with payment to income (PTI) between 40% and 50% | 29/08/2013 | 01/09/2013 | 01/09/2013 | HSE |
| 11 | The maximum variable rate portion of a mortgage loan cannot exceed 2/3, and the maximum portion of a variable rate that can change within 5 years from the date of approval cannot exceed 1/3. | 29/08/2013 | 01/09/2013 | 01/09/2013 | HSE |
| 12 | A loan shall not be approved or granted if the term to final repayment exceeds 30 years. | 29/08/2013 | 01/09/2013 | 01/09/2013 | HSE |

| | Stability Measures | Decision Date | Application Date | Influence Date | Type |
|---|---|---|---|---|---|
| 13 | A. Banking corporations should increase their Common Equity Tier I Capital target by a rate that represents 1 percent of their outstanding housing loans.<br>B. Banking corporations may reduce the risk weight attributed to variable interest leveraged loans from 100% to 75%.<br><br>**As the first decision was more influential than the second, ultimately the decisions were declared to be a stringency.** | 28/09/2014 | 01/01/2015 | 28/09/2014 | HSE |
| 14 | Basel III - Small & Large Banks: Tier 1 Capital target | 28/03/2012 | 01/01/2015 | 28/03/2012 | GEN |
| 15 | Basel III - Small & Large Banks: Total Capital Target | 28/03/2012 | 01/01/2015 | 28/03/2012 | GEN |
| 16 | Limitations on borrowers'/groups of borrowers' indebtedness: (1) the indebtedness of a borrower other than a bank to a banking corporation shall not exceed 15% of the banking corporation's capital; (2) the indebtedness of a group of borrowers to a banking corporation shall not exceed 25% of the banking corporation's capital; (3) the indebtedness of a banking group of borrowers to a banking corporation shall not exceed 15% of the banking corporation's capital; (4) the indebtedness of a controlled group of borrowers to a banking corporation shall not exceed 50% of the banking corporation's capital; (5) the total indebtedness of all "borrowers," "groups of borrowers," and "banking groups of borrowers," whose net indebtedness to the banking corporation exceeds 10% of the banking corporation's capital, shall not exceed 120% of the banking corporation's capital. | 09/06/2015 | 01/01/2016 | 09/06/2015 | CON |
| 17 | Basel III - Large Banks | 28/03/2012 | 01/01/2017 | 01/01/2015 | GEN |

*Notes*: Housing combines two categories: Real Estate Credit (REC) and LTV limits. CON stands for concentration limits; HSE stands for housing macroprudential regulation; GEN stands for general capital requirements. Application Date is the date of the MaP measure's application, Decision Date is the communication's date, and Influence Date is the date since the MaP measures were assumed to influence the credit aggregates.

In addition to the MaP measures that the Bank of Israel implemented, other factors that were no less important affected both the supply side, such as government measures to increase the supply of land available for construction and to increase the number of housing starts, and the demand side, such as government fiscal measures (taxation).

### 2.5.2 General Capital Requirement Measures

Another group of MaP measures, based on capital tools, aims to increase the banking system's stability and reduce systemic risks. Until 2010 capital requirements were only a recommendation to the banks. In mid-2010, the Bank of Israel's Banking Supervision Department started implementing the Basel II framework and set the core capital ratio target to a minimum of 7.5 percent. The banks had to achieve the Banking Supervision requirement by the end of 2010. Naturally, achieving the minimum capital ratio target on time forced the banks to adjust their core capital ratio right after the requirement was published.

The implementation of Basel III in Israel led the Banking Supervision in 2012 to announce an update of the core capital ratio requirement to no lower than 9 percent by the beginning of 2015. Two banks (out of the seven in our sample), classified as large banks, were required to achieve a core capital ratio of no lower than 10 percent by the beginning of 2017. As before, the adjustment processes started right after the publication of the requirements. Concerning the excess requirement of the large banks, we assume that the adjustment processes to achieve the extra 1 percent core capital ratio started right after the general requirement of 9 percent for all seven banks had to be fulfilled, i.e., the beginning of 2015.

As part of the Basel III framework, the Banking Supervision Department implemented a minimum liquidity coverage ratio of 100 percent. Liquidity requirements limit the credit supply so that banking corporations have to increase the liquidity assets in their balance sheets. Moreover, an additional leverage ratio surcharge also helps to control the banking system's leverage. It may influence the banks' capital and, as a result, credit supply. In our analysis, we refer to liquidity and capital tools together as bank stability tools.

### 2.5.3 Concentration Measures

Our analysis distinguishes between MaP measures targeting lending concentration, i.e., measures limiting credit to big borrowers, and general capital requirement measures, since each of the two groups may affect banks' credit differently. As a result of the big borrowers' dominance in the business sector, a MaP directive limiting the indebtedness of borrowers or groups of borrowers was announced in 2003. This MaP measure limited the banks' capital exposures to big borrowers, thus changing the banking sector supply for credit distribution by increasing the proportions of small and medium businesses and households. In 2015, more restrictive measures for big borrowers and groups of borrowers were imposed.

## 3 The Data

### 3.1 Bank Credit

The bank credit data are taken from the domestic banks' confidential reports to the Bank of Israel, submitted to the Banking Supervision Department on a quarterly basis in accordance with Banking Supervision Directive 831.

The data include credit to the public,[7] deposits, and other assets and liabilities, excluding bonds and securities that were borrowed or purchased under repurchase agreements. The credit data is disaggregated to commercial credit, which we refer to as business credit,[8] housing credit, and other credit to households (consumer credit).

Other than the credit data, the source for the rest of the banking system's data is the publicly available quarterly reports of the banks, describing the level of liquid assets, liabilities to assets, and deposits to assets. We include cash, deposits, and securities in the liquid assets category, provided that they are not encumbered to the lenders.

Our panel includes the seven largest banks in the domestic banking sector, which account for about 99 percent of the sector (see Section 2.3).

## 3.2   Capital Adequacy

Capital adequacy measures the amount of capital available to a bank to face unexpected losses that can be incurred due to a realization of risks to which the bank is exposed. The supervision policy is expressed in two ways: through a minimum capital requirement and by ensuring the quality of the capital. Besides applying the various Basel rules in Israel, the Bank of Israel's Banking Supervision Department establishes further requirements.

Since 2010, the Banking Supervision Department's requirements have dealt separately with two sets of capital: total capital (including tier one, tier two, and tier three capital) and tier one capital, or core capital, on its own—a distinction that preceded the final recommendation of Basel III.

In our empirical exercise, we use the gap between the ratio of actual capital to risk components of the banks and the Banking Supervision's most recent requirement, even if it had not yet been implemented. For example, in March 2012, the Banking Supervision published minimum core capital targets, which the banks had to reach by the beginning of 2015. In the paper, the gap for the years 2012–15 is calculated as the difference between the actual ratio for each period and the target ratio for 2015, with the understanding that the banks' policy changed during these years to reach the target.

The source of the capital adequacy data is the financial statements published by the banks each quarter, yet the requirement for "tier one" capital was declared only in 2010. Therefore, before 2010, the gap between the ratio of actual capital to risk components of the banks and the most recent Banking Supervision requirement refers to the "total capital" data that have been available since the 1990s.

## 3.3   Monetary Policy Surprises

As is customary in the literature (Gürkaynak et al., 2005; Gertler and Karadi, 2015; Cesa-Bianchi et al., 2020; Andrade and Ferroni, 2021), we use the unexpected component of the interest rates as an (exogenous) indicator of monetary policy to avoid endogeneity issues.

There are several possible empirical measurements of monetary policy surprises. The most common measurement, based on the financial markets' expectations, was first analyzed by Gürkaynak et al. (2005) for the US. Although Kutai (2020) suggests a similar methodology for Israel in line with the literature, due to data limitations, the period the Israeli data covers, starting from 2007, is too short for our estimations. Consequently, the monetary policy surprises we use in our estimations are derived from the difference between the interest rate set by the Bank of Israel and the average of all the available forecasts made by professional forecasters before the interest rate decision, which we extracted from Bloomberg, starting from 2003.

---

[7] During the first quarter of 2011, changes were made in the Banking Supervision Department directives. This created a break in the banking system's credit data. In order to deal with this discontinuity of the series, we included a dummy variable in the estimation to represent the reporting change.

[8] Including agriculture, industry, electricity, water, construction and real estate, commerce, transportation, information and telecommunications, and other business credit, excluding credit for financial services.

Following the GFC, we can observe several periods of monetary surprises captured by both methodologies. Between 2012 and 2014, some monetary policy decisions surprised both market forecasts and expert forecasts.

Figure 7 shows that the two measures have similar patterns, with a few exceptions, leading to a correlation coefficient of around 0.84. Our data for monetary policy surprises derived using expert forecasts span a more extended period than the data constructed using market-based forecasts. Thus, the correlation justifies our use of monetary policy surprises based on expert forecasts in our estimations.

We aggregate the surprises to quarterly data by summing surprises related to each decision during the corresponding quarter.[9] As shown in Figure 7, surprises are almost absent after mid-2015, in view of the stable interest rate of 0.1 percent since that time.

**Figure 7: Market- and Expert-Based Monetary Policy Surprises**

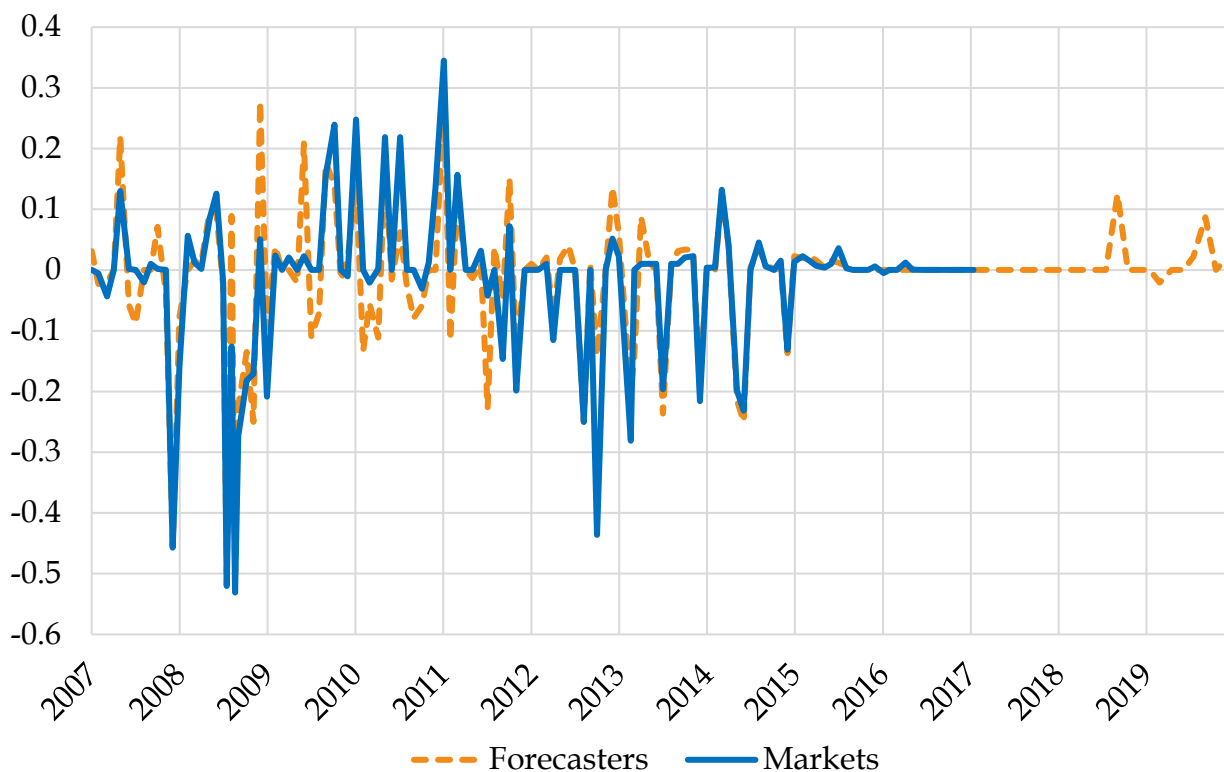

*Sources*: Kutai (2020) and Bloomberg.

Our estimation also includes monetary surprises in the US, and in an alternative specification, we add surprises in the eurozone (see the Appendix), both presented in Figure 8. As noted above, the Israeli banking system is entirely local, with practically no active foreign-owned banks. Moreover, the share of loans denominated in foreign currency is relatively small, especially in the consumer and mortgage sectors. Still, as the ability to substitute local currency credit with credit denominated in (or linked to) foreign currency exists, we opt to test the importance of this factor on the bank credit growth rate.

---
[9] Until April 2017, interest rate decisions were made by the Bank of Israel 12 times a year, toward the end of each month. Since April 2017, the Committee has decided on the interest rate eight times a year.

**Figure 8: Monetary Policy Surprises in the US, Eurozone, and Israel**

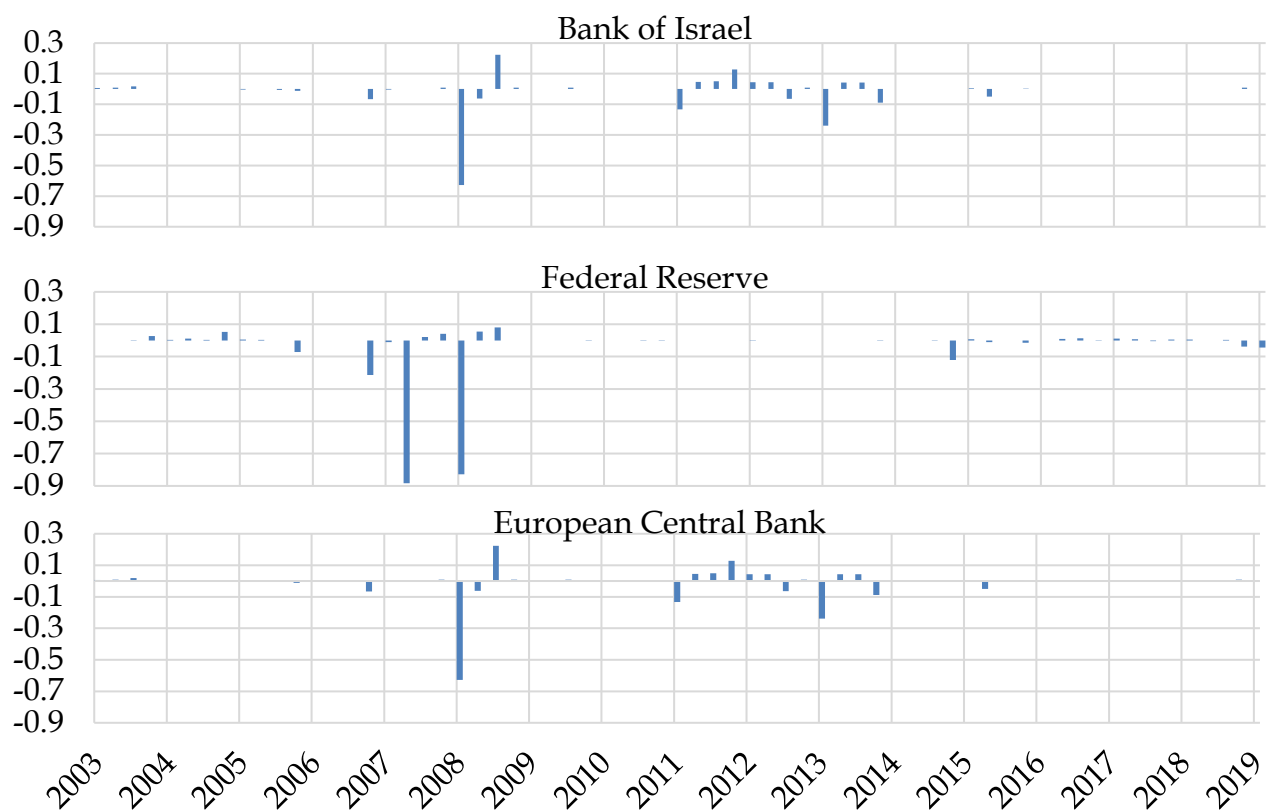

*Source*: Bloomberg.

We derive monetary policy surprises in the US and eurozone economies using the same methodology as for the Israeli economy (Figure 8). As seen in Figure 8, the magnitude of the ECB's monetary policy surprises is much smaller than that of the Fed. The correlation between these measured surprises is about 0.5 between any two measures.

### 3.4   Prudential Measures

We refer in our estimation to the list of MaP measures summarized in Table 1. We classify the MaP measures into three groups:[10] MaP measures on the **housing** market; measures related to limitations on core capital level, which refer to the **general** bank credit market; and measures related to credit **concentration**. As shown in Table 1, most of the MaP measures target the housing sector. We cannot identify the importance of each of these measures ex ante. Therefore, as is well accepted in this literature (e.g., Cerutti et al., 2017), we assume that each of the measures has the same weight, gather all the measures into one group, and construct a cumulative index of these measures.

---

[10] Our classification of all MaP measures is based on the IBRN international dataset of MaP measures.

The first group includes measures aimed at the housing market, and can be characterized as measures limiting borrowers, such as LTV (loan-to-value) and PTI (payment-to-income) limits, or as measures that set limitations on the banks, such as differential or excess risk weights.

The second group includes general capital and liquidity requirement measures, i.e., measures mainly referring to the implementation of Basel capital agreements (hereinafter: "general"); the third group includes measures aimed at lessening banks' credit concentration by forcing limitations on credit to big borrowers. Even though the latter two may be related to the bank's business credit sector, these groups are separated in the benchmark estimation since we want to test whether each of these two groups affected housing and consumer credit differently.[11] We exclude two types of measures related to reserve requirements on foreign currency-denominated accounts and interbank exposure limits. These types include only three MaP measures that aimed to address specific issues and were not supposed to affect the credit sector we investigated.[12]

General MaP measures were aimed at increasing the bank's soundness, which can affect all bank credit sectors or a specific sector. Reducing the growth rate of business credit may be the preferred reaction to the general MaP measures, due to their high weight as risk assets. The concentration MaP measures are aimed at reducing the share of big borrowers in each bank's assets, i.e., reducing the bank's business credit volume.

In the estimation of each of the types of credit—housing, consumer, business, and total credit—we include the three groups of MaP measures. We expect measures that refer to the studied credit sector to moderate its growth rate, while MaP measures that refer to other types of credit, which are substitutes for the banks as suppliers of credit, or for the borrowers as demanders of credit, may work to increase the growth rate of the credit studied.

Each of the MaP categories is assigned the value of 1 in periods in which at least one MaP measure in the category became effective (according to the column "Influence Date" in Table 1). We allow different time spans for MaP measures to affect the market. For housing MaP measures, we include the accumulation of measures since the beginning of the studied period as the indicator variable, lagged one period, as we assume that the implemented measures continue to affect the market as long as they are in place and reflect the MaP policy stance. General MaP measures are included in the estimation accumulated over eight quarters, lagged one-quarter relative to the date they became effective. We use this accumulated period since after the threshold for capital adequacy is achieved (plus some buffer), the effect of the restriction on the credit growth rate ceases. Eight quarters is a period that covers the maximum number of periods between the time a general MaP measure began influencing banks' behavior and the time it had to be applied (See Table 1). All the thresholds were achieved by the banks before the due date, and so accumulating over a longer period is unnecessary. Once banks achieve the recommended capital adequacy, the MaP policy stance toward tightening banking risks is fulfilled.

The concentration measures were accumulated over four quarters, lagged one period. We accumulate four quarters backward since the volume of business credit that a bank extends, on which these measures are intended to operate, is approved and validated once a year. Implementation of the concentration restrictions may therefore be fulfilled over one year after the credit portfolio of all businesses come up for reapproval. For the estimation of the business credit sector, we lag the accumulated measures by four quarters, rather than by one quarter as in the other sectors. This is because while shifting credit from the business sector to the other sectors can lead this process, the process of reducing the credit of the business sector can be done only after the reapproval of a credit portfolio under the new concentration restrictions.

---

[11] Our robustness checks aggregate the general and concentration groups discussed in Section 5.

[12] In 2011, the Bank of Israel changed definitions related to interbank loans, which we exclude from our data (we use credit for the nonfinancial sector only), and raised the reserve requirement imposed on banking corporations for FX derivative transactions by nonresidents (we use local credit only).

Table 2 reveals that, with the exception of one MaP measure, all of the MaP measures were implemented after the GFC. Therefore, there is no need to distinguish in the estimations between pre- and post-GFC when relating to the MaP measures or their interaction with monetary policy. It is also evident that the use of housing MaP measures is more intensive than other measures due to the rapid increases in house prices during these years (Figures 3 and 4).

**Table 2: Macroprudential Measures by Category and Quarter Imposed**

|        | Concentration | General | Housing |
|--------|---------------|---------|---------|
| Sep-03 | 1             |         |         |
| Mar-10 |               |         | 1       |
| Jun-10 |               | 1       |         |
| Sep-10 |               |         | 1       |
| Dec-10 |               |         | 1       |
| Jun-11 |               |         | 1       |
| Jun-12 |               | 1       |         |
| Dec-12 |               |         | 1       |
| Mar-13 |               |         | 1       |
| Sep-13 |               |         | 1       |
| Sep-14 |               |         | 1       |
| Mar-15 |               | 1*      |         |
| Jun-15 | 1             |         |         |

*Notes*: *Large banks only.

### 3.5 Other Macroeconomic Variables

In order to take into account the effect of the business cycle and housing market activity on credit aggregates, we add control variables such as unemployment rate, change in real wages, business sector GDP growth, and real change in house prices. A detailed description of the macroeconomic variables in each specification is presented below in Section 4.

### 3.6 Sample Period

The estimation is carried out as a weighted cross-section panel estimation, correcting for cross-section heteroscedasticity and contemporaneous correlation. The estimation is performed for seven banks over 63 periods (2004:Q2–2019:Q4) for a total of 441 observations. The dependent variable is the log difference of the real quarterly volume of credit.

Due to the substantial impact of the GFC on policy and the market, we distinguish monetary policy surprises before and after the GFC, starting from the third quarter of 2008.

## 4 Estimation and Results

As described in Section 2, the banking sector in Israel is characterized by a small number of domestic banks for which lending activity is concentrated in the domestic market, and its sources are domestic as well. That is, the lending activity abroad of the domestic banks and the activities of foreign banks in the domestic market are marginal. Naturally, those qualities lead us to focus on estimates of the influence of the domestic monetary policy and domestic MaP measures, and the interaction between them, on the domestic banks' lending. Because Israel is a small open economy, we complement the specification with monetary policy surprises from the US,[13] as described in Section 3.2, in order to allow us to test the effect of the interest rate differential between domestic and foreign financing on the growth of domestic credit.

Our first objective is to assess the effect of domestic MaP policies—including their interaction with monetary policy—on the *housing* credit (mortgage) market, which is one of the components of the domestic banks' total lending activity.

The second component of bank credit is lending to the *business* sector (excluding the financial sector). We assess the effect of domestic MaP policies and of the interactions of these policies with monetary policy on business sector lending.

Nonhousing *consumer* credit is the third component of bank credit. This credit is not backed by the borrower's housing, and is usually used for consumption, but may also serve—for the banks as credit suppliers, and for households as credit demanders—as a substitute for mortgages. As a final step, we assess the effect of monetary and MaP policies on total bank credit, which is comprised of the above three components.

The general formulation of our estimation is

$$\Delta Credit_{i,b,t} = a_i + \beta_{1i}MaP_t + \beta_{2i}MonPol_t + \beta_{3i}MaP_t \times MonPol_t + \beta_{4i}ForeignMonPol_t + \beta_{5i}Macro_{i,t} + \beta_{6i}BankChar_{i,b,t} + \epsilon_{i,b,t},$$

where $t$ spans from 2004:Q2 to 2019:Q4, and $b$ stands for the bank. $\Delta Credit_{i,b,t}$ is the log change of domestic bank credit of type $i$ (housing, consumer, business, or total credit) by bank $b$ in quarter $t$. $MaP_t$ are indices of MaP measures and $MonPol_t$ and $ForeignMonPol_t$ are the unexpected change in the Bank of Israel and foreign interest rates. The estimated parameters $a$ and $\beta$, macroeconomic control variables ($Macro_{i,t}$), and bank-specific control variables ($BankChar_{i,b,t}$) vary by type of credit $i$. We will elaborate on the specific variables and lags included in the estimations in the sections describing the results. $\epsilon_{i,b,t}$ are cross-section-weighted and allow for conditional cross-section correlation between the contemporaneous residuals but are restricted to be uncorrelated between different periods. The estimation results are presented in Tables 3 to 6.

We choose to separately analyze the components of local bank credit—housing, business, and consumer credit—in order to study the effect of each of the three classes of MaP measures on each credit type. While measures targeting housing credit were meant to moderate its growth in order to minimize the risk to the banking sector's stability, they may have induced the substitution of housing credit with other types of credit (such as business credit), either due to a shift in the demand for this credit due to limitations on borrowers, or due to supply-side considerations on the part of the banks, so that the effect of these measures on total bank credit may have been small or insignificant. The same may hold for the general and concentration MaP measures: they may have induced a possible indirect substitution effect alongside the direct moderating effect on business, and possibly consumer and housing, credit. In order to analyze the direct and indirect effects of the three classes of MaP measures, we choose to include all three in the estimation of each of the separate credit components. In addition, we estimate their effect on the growth rate of total domestic bank credit, which comprises housing, consumer, and business credit.

---

[13] In an alternative specification, we also include monetary policy surprises from the eurozone. The qualitative results remain unchanged (see the Appendix).

As mentioned above, we include in the estimations each of the three MaP housing, concentration, and general measures accumulated over different lengths of periods. While we find that the effect of housing MaP measures on the mortgage sector is best exhibited when considering the possible effect of accumulated measures in the past on this market, we find that the general and concentration MaP measures affect credit dynamics when we consider the accumulation of the MaP measures over a shorter period into account (Figure 9).

**Figure 9: Cumulative MaP Measures**

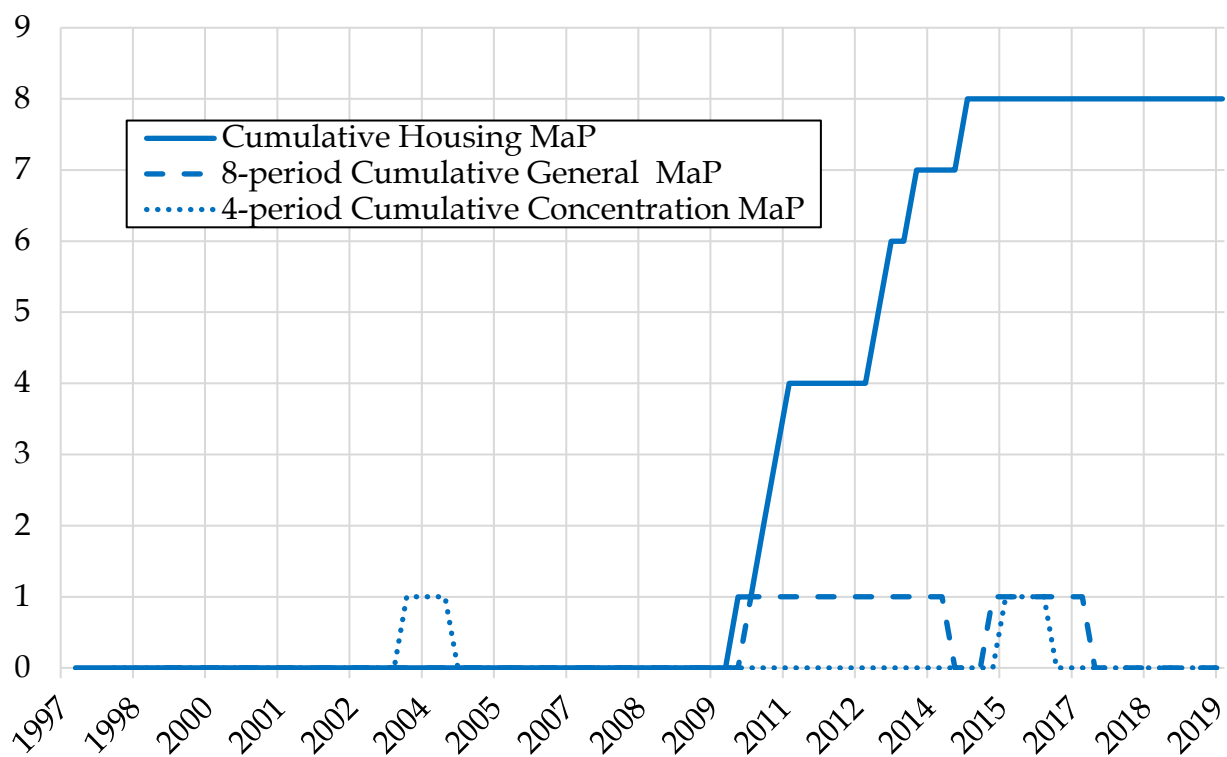

*Notes*: MaP dummy variables were used in the estimations. *Source*: Bank of Israel Research Department.

The interaction between MaP measures and monetary policy is specified as the product of the MaP measures index of dummies and the lagged one- and two-period monetary policy. Foreign monetary policy is included in the estimations with one and two lags.

For each of the credit types, we present four specifications. Column 1 of each of the tables includes the basic specification. Column 3 presents the specification including global monetary policy surprises, and Columns 2 and 4 add the interaction terms between MaP measures and monetary policy to these specifications. In the specifications shown in Column 4 of Tables 3 to 6, we represent global monetary policy by including monetary policy shocks in the US, as most of the foreign currency credit is denominated in dollars. We present an alternative specification including eurozone monetary policy surprises in the Appendix. Our qualitative results remain robust to this alternative specification.

In the following estimation results, the significance of the sum of the monetary policy lags coefficients and the interaction lags coefficients is determined using the Wald test.

## 4.1 Housing Credit Estimation Results

Table 3 shows that MaP measures relating to the housing market succeeded in reducing the growth rate of housing credit according to all specifications. General MaP measures contributed to reducing the growth rate of housing credit, although the results were significant only in the specifications excluding the interaction terms.

**Table 3: Housing Credit, Sample 2004:Q2 – 2019:Q4, 7 Banks**

| | | 1<br>Basic – excl. interactions and global policy | 2<br>Including interactions | 3<br>Including global policy | 4<br>Including interactions and global monetary policy |
|---|---|---|---|---|---|
| MaP | Cumulative MaP_housing(-1), since 2004 | -0.001** | -0.002** | -0.002*** | -0.002*** |
| | MaP_general(-1) to (-8) | -0.004* | -0.002 | -0.005* | -0.004 |
| | MaP_concentration(-1) to (-4) | 0.002 | 0.006 | 0.007* | 0.010** |
| Monetary Policy | Monetary policy(-1) until the GFC | -1.89* | -2.23** | -1.34 | -2.08 |
| | Monetary policy(-2) until the GFC | -4.56*** | -4.53*** | -2.67** | -2.78** |
| | **Aggregated before 2008** | **-6.45*** | **-6.76*** | **-4.00** | **-4.86*** |
| | Monetary policy(-1) after the GFC | 0.82 | 0.88 | -0.41 | -2.70 |
| | Monetary policy(-2) after the GFC | 0.75 | 0.56 | 0.22 | -0.29 |
| | **Aggregated after 2008** | **1.57*** | **1.43** | **-0.20** | **-2.99** |
| Interactions | Cumulative MaP_housing(-1) × Mon. policy(-1) | | -0.14 | | 0.24 |
| | Cumulative MaP_housing(-1) × Mon. policy(-2) | | -0.22 | | -0.15 |
| | **Aggregated MaP_housing × Mon. policy** | | **-0.36** | | **0.09** |
| | MaP_general(-1) to (-8) × Mon. policy(-1) | | 1.04 | | 2.38 |
| | MaP_general(-1) to (-8) × Mon. policy(-2) | | 2.76 | | 2.83 |
| | **Aggregated MaP_general × Mon. policy** | | **3.80** | | **5.21*** |
| | MaP_concentr.(-1) to (-4) × Mon. policy(-1) | | 4.08 | | 3.90 |
| | MaP_concentr.(-1) to (-4) × Mon. policy(-2) | | 3.09 | | 2.94 |
| | **Aggregated MaP_concentr. × Mon. policy** | | **7.18*** | | **6.85*** |
| Foreign | US monetary policy(-1) until the GFC | | | -1.89 | -1.27 |
| | US monetary policy(-2) until the GFC | | | -3.00** | -2.93** |
| | **US Aggregated before 2008** | | | **-4.88*** | **-4.20** |
| | US monetary policy(-1) after the GFC | | | 2.17 | 5.64* |
| | US monetary policy(-2) after the GFC | | | 0.78 | 2.04 |
| | **US Aggregated after 2008** | | | **2.95** | **7.68** |
| Banks^ | Excess Tier(-1) | 0.38*** | 0.40*** | 0.57*** | 0.57*** |
| Macro | D(real wage(-2)) | 0.12* | 0.14** | 0.18** | 0.19** |
| | D(unemployed 25-64(-5)) | -0.77*** | -0.61** | -0.71** | -0.58* |
| | D(log(house_prices(-1))) | 0.07*** | 0.06** | 0.05* | 0.05 |
| | Seasonal Dummies | Yes | Yes | Yes | Yes |
| | Data Break Dummies | Yes | Yes | Yes | Yes |
| | C | 0.001 | 0.001 | -0.004 | -0.003 |
| | CRISIS2008 | 0.01** | 0.02*** | 0.02*** | 0.02*** |
| | **Observations** | 441 | 441 | 441 | 441 |
| | **Adjusted R²** | 0.17 | 0.17 | 0.18 | 0.18 |

*Notes*: ***, **, and * indicate significance at the 1%, 5%, and 10% level, respectively. ^ The estimation also includes the change from liabilities to assets, the change in the banks' real assets, the change in deposits as a share of assets, the change in liquid assets, and the change from excess reserves to assets, with varying lags. All results were insignificant.

We also find that MaP measures targeting the concentration of the banks' activity increased housing credit growth (the results are significant only in the full specification, Column 4). This demonstrates a substitution effect on business credit from the supply side of the banks, increasing mortgage lending as a substitute for business lending, particularly to large borrowers. We find that positive monetary policy surprises (unexpected tightening) significantly decreased the growth rate of housing credit, but turned insignificant in the period after the financial crisis, starting in the third quarter of 2008 (Table 3). For some of the specifications, the interaction between the general or concentration MaP measures and monetary policy amplifies the substitution effect on mortgage credit.

**Figure 10: Housing Credit Decomposition**

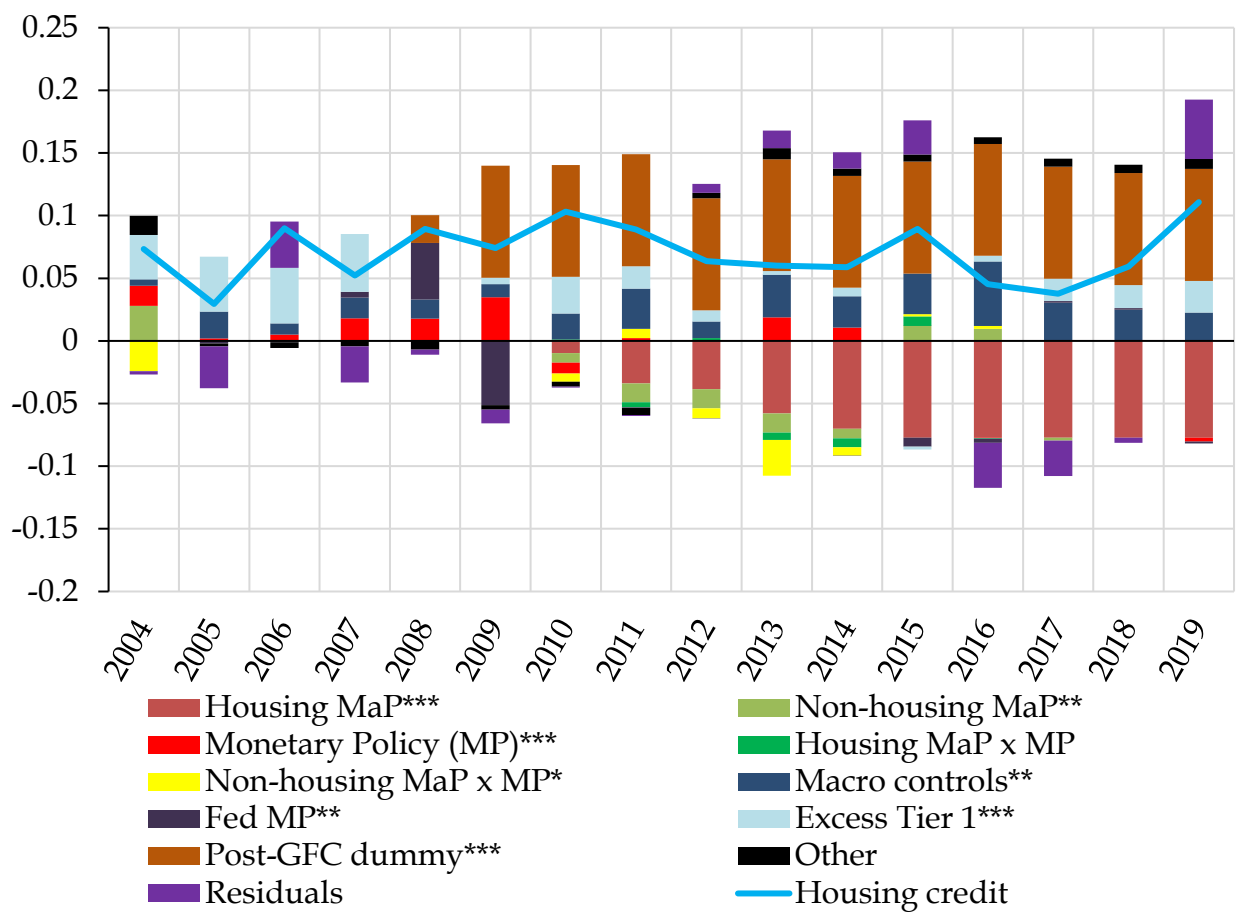

*Notes*: Housing credit is expressed in log difference (dlog). "Other" includes the intercept, bank fixed effects, dummies to control for seasonality, data break dummies, and other financial ratios. ***, **, and * indicate significance at 1%, 5%, and 10% level, respectively. *Source*: Estimation, Table 3, Column 4.

As Israel is a small open economy, we look for the effect of global interest rates, represented by US monetary policy. We find that higher rates abroad tended to moderate local housing credit growth before the GFC, rather than increasing it, as would have been if local housing debt and foreign debt were substitutes. Given that the Israeli mortgage market is (almost) entirely a local market, the negative effect on the growth rate of local mortgages may indicate that US surprises are a proxy for expected domestic monetary policy, which slowed the growth rate of mortgages before the GFC. Another possible interpretation links the higher foreign interest rate to the actual and expected local currency appreciation. Since house prices were generally denominated in US dollars before the GFC, an appreciation would lead, at least in the short term, to a moderation of the local currency value of house prices and, therefore, of mortgages.

The macroeconomic indicators—real wages, unemployment, and house prices—all have the expected effect on the mortgage market. Higher prices indicate excess demand in the market, but also increase the nominal value of a given number of transactions, inducing higher demand for mortgages.

We also control for bank characteristics that may affect the banks' tendency to offer credit. An increase in the banks' capital relative to the supervisory requirements (lagged two periods), i.e., a positive gap, tends to increase the growth of housing credit.

Figure 10 presents the average annual contributions of the various factors that affect the dynamics of housing credit. Housing MaP measures made a substantial downward contribution to the development of housing credit after 2011. Foreign monetary policy had some impact on local housing credit in the years before the GFC, and did not contribute to credit dynamics in the years after the GFC. Macroeconomic variables and bank characteristics contributed to the housing credit increase after 2009.

The post-GFC dummy reflects post-2008 factors, such as unconventional monetary policies, that sustained the demand for housing credit. Figure 10 also shows that excess tier 1 capital increased housing credit.

## 4.2 Business Credit Estimation Results

Although the banks' share of total business credit declined in the past decade from about 75 percent to 50 percent (Figure 5b), business credit is still an important sector, accounting for about a 40 percent share of total bank credit.

We find that general MaP measures tended to reduce the growth rate of business credit, as expected, although the results were significant only in some of the specifications (Table 4). In addition, our results show that MaP measures aimed at housing credit tended to increase the supply of credit to the business sector, as those are substitutes on the asset side of the banks' balance sheets.

We do not find evidence that the concentration MaP measures, which were implemented only once during our sample period, had any significant effect on business credit.[14] Similar to our results for housing credit, higher policy rates tended to reduce business credit growth before the GFC, but this influence diminishes and we find that higher policy rates tend to increase business credit after the GFC, contrary to our basic intuition. We find that the interactions between monetary policy and the MaP measures under the full specification do not significantly affect business credit.

There is partial evidence that less accommodative foreign monetary policy reduced the growth rate of domestic business credit after 2008. This suggests that the signaling effect of a surprise in foreign interest rates, assuming one exists, is stronger than the possible substitution effect on foreign credit. Similar to our results for housing credit, macroeconomic indicators and bank-specific indicators have the expected effect on the evolution of the business credit aggregates (Table 4).

---

[14] We included one measure taken in September 2003, which is before the sample period but, due to lags in the specification, is also relevant.

# Table 4: Business Credit, Sample 2004:Q2 – 2019:Q4, 7 Banks

| | | 1<br>Basic – excl. interactions and global policy | 2<br>Including interactions | 3<br>Including global policy | 4<br>Including interactions and global monetary policy |
|---|---|---|---|---|---|
| MaP | Cumulative MaP_housing(-1), since 2004 | 0.001** | 0.001** | 0.001*** | 0.001** |
| | MaP_general(-1) to (-8) | -0.008*** | -0.005* | -0.007** | -0.005 |
| | MaP_concentration(-4) to (-7) | -0.004 | -0.006 | -0.005 | -0.005 |
| Monetary Policy | Monetary policy(-1) until the GFC | 0.12 | -0.44 | 1.29 | 0.53 |
| | Monetary policy(-2) until the GFC | -2.75** | -2.17* | -3.71*** | -3.00* |
| | **Aggregated before 2008** | **-2.63*** | **-2.61** | **-2.42** | **-2.47** |
| | Monetary policy(-1) after the GFC | 1.28** | 0.22 | 2.97*** | 2.36 |
| | Monetary policy(-2) after the GFC | 3.32*** | 3.80*** | 3.46*** | 6.74*** |
| | **Aggregated after 2008** | **4.60*** | **4.02*** | **6.43*** | **9.10*** |
| Interactions | Cumulative MaP_housing(-1) × Mon. policy(-1) | | 0.23 | | -0.04 |
| | Cumulative MaP_housing(-1) × Mon. policy(-2) | | -0.43* | | -0.81** |
| | **Aggregated MaP_housing × Mon. policy** | | **-0.20** | | **-0.86** |
| | MaP_general(-1) to (-8) × Mon. policy(-1) | | 3.00 | | 2.25 |
| | MaP_general(-1) to (-8) × Mon. policy(-2) | | 3.57* | | 2.61 |
| | **Aggregated MaP_general × Mon. policy** | | **6.57*** | | **4.86** |
| | MaP_concentr.(-4) to (-7) × Mon. policy(-1) | | 1.89 | | 1.13 |
| | MaP_concentr.(-4) to (-7) × Mon. policy(-2) | | -2.66 | | -1.03 |
| | **Aggregated MaP_concentr. × Mon. policy** | | **-0.77** | | **0.1** |
| Foreign | US monetary policy(-1) until the GFC | | | -1.97 | -1.42 |
| | US monetary policy(-2) until the GFC | | | 1.54 | 1.03 |
| | **US Aggregated before 2008** | | | **-0.43** | **-0.39** |
| | US monetary policy(-1) after the GFC | | | -4.43** | -3.14 |
| | US monetary policy(-2) after the GFC | | | -0.60 | -5.40 |
| | **US Aggregated after 2008** | | | **-5.02** | **-8.54** |
| Banks^ | Excess Tier(-5) | 0.40*** | 0.29* | 0.31* | 0.36** |
| | D(Liabilities(-1)/Assets(-1)) | -0.73 | -1.06* | -0.88 | -1.01* |
| Macro | D(BusinessGDP(-2)) | 0.22*** | 0.20*** | 0.21*** | 0.21** |
| | D(unemployed 25-64(-2)) | -1.55*** | -1.77*** | -1.74*** | -1.63*** |
| | Seasonal Dummies | Yes | Yes | Yes | Yes |
| | Data Break Dummies | Yes | Yes | Yes | Yes |
| | C | -0.03** | -0.02** | -0.03*** | -0.02*** |
| | CRISIS2008 | 0.02*** | 0.01*** | 0.01** | 0.01*** |
| | **Observations** | 441 | 441 | 441 | 441 |
| | **Adjusted R²** | 0.30 | 0.30 | 0.30 | 0.30 |

*Notes*: ***, **, and * indicate significance at the 1%, 5%, and 10% level, respectively. ^ The estimation also includes the change from liabilities to assets, the change in the banks' real assets, the change in deposits as a share of assets, the change in liquid assets, and the change from excess reserves to assets, with varying lags. All results were insignificant.

## 4.3 Consumer Credit Estimation Results

As stated above, consumer credit comprises a relatively small share of the banks' total credit, and this share has been relatively constant during the past two decades. Therefore, our ability to account for changes in consumer credit due to changes in policy or macroeconomic factors is limited.

## Table 5: Consumer Credit, Sample 2004:Q2 – 2019:Q4, 7 Banks

|  |  | 1<br>Basic – excl. interactions and global policy | 2<br>Including interactions | 3<br>Including global policy | 4<br>Including interactions and global monetary policy |
|---|---|---|---|---|---|
| MaP | Cumulative MaP_housing(-1), since 2004 | -0.001 | -0.002 | -0.002 | -0.003 |
|  | MaP_general(-1) to (-8) | 0.003 | 0.005 | 0.000 | 0.005 |
|  | MaP_concentration(-1) to (-4) | -0.003 | 0.013 | 0.003 | 0.013 |
| Monetary Policy | Monetary policy(-1) until the GFC | -0.52 | -1.60 | 8.17* | 7.03 |
|  | Monetary policy(-2) until the GFC | -12.37*** | -12.73*** | -16.86*** | -17.77*** |
|  | **Aggregated before 2008** | **-12.89\*\*** | **-14.33\*\*** | **-8.68** | **-10.74** |
|  | Monetary policy(-1) after the GFC | -0.07 | -0.77 | -0.53 | -1.58 |
|  | Monetary policy(-2) after the GFC | 0.54 | 3.07 | -3.23 | 6.83 |
|  | **Aggregated after 2008** | **0.47** | **2.30** | **-3.76** | **5.25** |
| Interactions | Cumulative MaP_housing(-1) × Mon. policy(-1) |  | -0.80 |  | -.95 |
|  | Cumulative MaP_housing(-1) × Mon. policy(-2) |  | -2.05** |  | -2.38* |
|  | **Aggregated MaP_housing × Mon. policy** |  | **-2.84\*\*** |  | **-3.33\*** |
|  | MaP_general(-1) to (-8) × Mon. policy(-1) |  | 6.2 |  | 9.55 |
|  | MaP_general(-1) to (-8) × Mon. policy(-2) |  | 7.21 |  | 4.89 |
|  | **Aggregated MaP_general × Mon. policy** |  | **13.41** |  | **14.44** |
|  | MaP_concentr.(-1) to (-4) × Mon. policy(-1) |  | 7.88 |  | 0.40 |
|  | MaP_concentr.(-1) to (-4) × Mon. policy(-2) |  | 19.06 |  | 18.93 |
|  | **Aggregated MaP_concentr. × Mon. policy** |  | **26.94\*** |  | **19.33** |
| Foreign | US monetary policy(-1) until the GFC |  |  | -11.84*** | -11.10** |
|  | US monetary policy(-2) until the GFC |  |  | 5.67 | 6.64 |
|  | **US Aggregated before 2008** |  |  | **-6.17** | **-4.46** |
|  | US monetary policy(-1) after the GFC |  |  | -1.26 | 1.57 |
|  | US monetary policy(-2) after the GFC |  |  | 7.73 | -6.50 |
|  | **US Aggregated after 2008** |  |  | **6.46** | **-4.92** |
| Banks^ | Excess Tier(-1) | -0.15 | 0.04 | -0.01 | 0.05 |
|  | D(Liabilities(-1)/Assets(-1)) | -3.98** | -4.29** | -4.45** | -4.49** |
| Macro | D(real wage(-1)) | 0.67*** | 0.62** | 1.00*** | 0.87*** |
|  | D(unemployed 25-64(-2)) | -0.94 | -0.37 | -1.87 | -1.22 |
|  | Seasonal Dummies | Yes | Yes | Yes | Yes |
|  | Data Break Dummies | Yes | Yes | Yes | Yes |
|  | C | 0.03** | 0.03* | 0.02 | 0.02 |
|  | CRISIS2008 | -0.005 | -0.003 | 0.007 | 0.004 |
|  | **Observations** | 441 | 441 | 441 | 441 |
|  | **Adjusted R²** | 0.04 | 0.04 | 0.05 | 0.05 |

*Notes*: ***, **, and * indicate significance at the 1%, 5%, and 10% level, respectively. ^ The estimation also includes the change from liabilities to assets, the change in the banks' real assets, the change in deposits as a share of assets, the change in liquid assets, and the change from excess reserves to assets, with varying lags. All results were insignificant.

The share of variance explained in our estimation is close to zero. Therefore, any conclusions referring to the drivers of consumer credit are limited. Our results are presented in Table 5 and show that we do not find any significant effect of MaP measures on consumer credit. Less accommodative monetary policy reduced the growth rate of consumer credit only before the GFC, similar to the effect found for business and housing credit.

The expected effect of MaP measures aimed at housing credit on consumer credit is ambiguous. MaP measures limiting housing credit or making it more expensive are expected to increase consumer credit due to substitution effects on both the demand and supply sides. By contrast, if housing (mortgage) credit and consumer credit are complements, limiting mortgages will also tend to reduce demand for consumer credit.

Looking at the interaction between monetary policy and housing market MaP measures, we find evidence for housing and consumer credit being substitutes. The interaction term of these measures with monetary policy is negative, indicating, as for business credit, that when stricter housing MaP measures accompany accommodative monetary policy, consumer (nonhousing) credit will tend to increase. On the other hand, coupling a surprising interest rate hike with accumulated housing market MaP measures intensifies the effect of monetary surprises on consumer credit.

We find that until 2008 a positive surprise to the global interest rate tended to decrease the growth rate of consumer credit. After 2008, this effect is not significant. As for housing credit, interest rate surprises abroad may be an indication of the expected local monetary policy and therefore affect the growth rate of consumer credit.

## 4.4 Total Credit Estimation Results

The previous sections analyzed the effect of MaP measures and monetary policy on each component of bank credit: housing, business, and consumer credit. Although housing MaP measures are naturally targeted at the housing market, the policy aims to influence the macroeconomic conditions and financial stability.

Therefore, it is important to examine the effect of these measures on total bank credit. We repeat our estimation for aggregate bank credit, which is the sum of the three components mentioned above. Table 6 presents the detailed results.

We find that MaP measures targeted at specific credit sectors, which influence these sectors, do not effect the dynamics of total credit according to our estimations. The effect of housing and general MaP measures on total credit is insignificant, and the accumulated effect of concentration-related measures is also null. As we found in the full specification (Column 4) for some of the specific credit sectors, monetary policy affected credit growth only before the GFC. A negative sign on the interaction between monetary policy and housing MaP measures in the restricted specification (Column 2) indicates that accommodative monetary policy tended to support total credit growth when coupled with restrictive housing MaP measures.

Figure 11 shows that monetary policy is not central in explaining total bank credit dynamics. The interaction between monetary policy and nonhousing MaP measures decreased total credit growth between 2012 and 2014.

The housing MaP measures made a significant contribution to changes in housing and business credit. However, the substitution between the types of credit offsets these effects, resulting in nearly no contribution to changes in total credit. As Figure 11 shows, this influence was augmented by the interaction of housing MaP measures and monetary policy, contributing to total credit mainly through the effect of the interaction on consumer credit.

# Table 6: Total Credit, Sample 2004:Q2—2019:Q4, 7 Banks

| | | 1<br>Basic – excl. interactions and global policy | 2<br>Including interactions | 3<br>Including global policy | 4<br>Including interactions and global monetary policy |
|---|---|---|---|---|---|
| MaP | Cumulative MaP_housing(-1), since 2004 | 0.0002 | -0.0002 | 0.0006 | -0.0000 |
| | MaP_general(-1) to (-8) | -0.0002 | 0.002 | 0.0001 | 0.002 |
| | MaP_concentration(-1) to (-4) | 0.006** | 0.008** | 0.008** | 0.010*** |
| | MaP_concentration(-4) to (-7) | -0.008*** | -0.009*** | -0.007*** | -0.008*** |
| Monetary Policy | Monetary policy(-1) until the GFC | -0.10 | -0.68 | 2.44** | 1.67 |
| | Monetary policy(-2) until the GFC | -4.60*** | -4.74*** | -5.82*** | -6.01*** |
| | **Aggregated before 2008** | **-4.70*** ** | **-5.42*** ** | **-3.38** ** | **-4.34*** ** |
| | Monetary policy(-1) after the GFC | 0.59 | 0.17 | 1.06 | -0.82 |
| | Monetary policy(-2) after the GFC | 1.76*** | 2.31*** | 1.27* | 2.44 |
| | **Aggregated after 2008** | **2.35*** ** | **2.48*** ** | **2.32*** | **1.63** |
| Interactions | Cumulative MaP_housing(-1) × Mon. policy(-1) | | -0.12 | | -0.02 |
| | Cumulative MaP_housing(-1) × Mon. policy(-2) | | -0.62*** | | -0.58** |
| | **Aggregated MaP_housing × Mon. policy** | | **-0.74*** | | **-0.60** |
| | MaP_general(-1) to (-8) × Mon. policy(-1) | | 2.69** | | 3.62** |
| | MaP_general(-1) to (-8) × Mon. policy(-2) | | 4.31*** | | 3.92** |
| | **Aggregated MaP_general × Mon. policy** | | **6.99*** ** | | **7.54*** ** |
| | MaP_concentr.(-1) to (-4) × Mon. policy(-1) | | 7.83*** | | 6.27*** |
| | MaP_concentr.(-1) to (-4) × Mon. policy(-2) | | -0.17 | | 0.50 |
| | **Aggregated MaP_concentr. × Mon. policy** | | **7.66*** | | **6.78*** |
| | MaP_concentr.(-4) to (-7) × Mon. policy(-1) | | -2.04 | | -2.59 |
| | MaP_concentr.(-4) to (-7) × Mon. policy(-2) | | 1.49 | | 2.16 |
| | **Aggregated MaP_concentr. × Mon. policy** | | **-0.55** | | **-0.43** |
| Foreign | US monetary policy(-1) until the GFC | | | -3.17*** | -2.47** |
| | US monetary policy(-2) until the GFC | | | 1.48 | 1.40 |
| | **US Aggregated before 2008** | | | **-1.68** | **-1.07** |
| | US monetary policy(-1) after the GFC | | | -1.38 | 1.65 |
| | US monetary policy(-2) after the GFC | | | 1.08 | -0.05 |
| | **US Aggregated after 2008** | | | **-0.30** | **1.60** |
| Banks^ | Excess Tier(-1) | 0.27*** | 0.32*** | 0.33*** | 0.35*** |
| | D(log(Real Assets(-1))) | 0.08 | 0.08* | 0.06 | 0.07 |
| Macro | YoY(Real Wage(-2)) | 0.01 | 0.04 | -0.10 | -0.04 |
| | D(Unemployed 25-64(-2)) | -1.15*** | -1.01*** | -1.42*** | -1.26*** |
| | YoY(House Prices(-1)) | 0.01 | 0.001 | -0.002 | -0.01 |
| | Seasonal Dummies | Yes | Yes | Yes | Yes |
| | Data Break Dummies | Yes | Yes | Yes | Yes |
| | C | -0.002 | -0.002 | -0.003 | -0.003 |
| | CRISIS2008 | 0.007 | 0.009* | 0.007 | 0.01* |
| | **Observations** | 441 | 441 | 441 | 441 |
| | **Adjusted R²** | 0.28 | 0.32 | 0.29 | 0.32 |

*Notes*: ***, **, and * indicate significance at the 1%, 5%, and 10% level, respectively. ^ The estimation also includes the change from liabilities to assets, the change in the banks' real assets, the change in deposits as a share of assets, the change in liquid assets, and the change from excess reserves to assets, with varying lags. All results were insignificant.

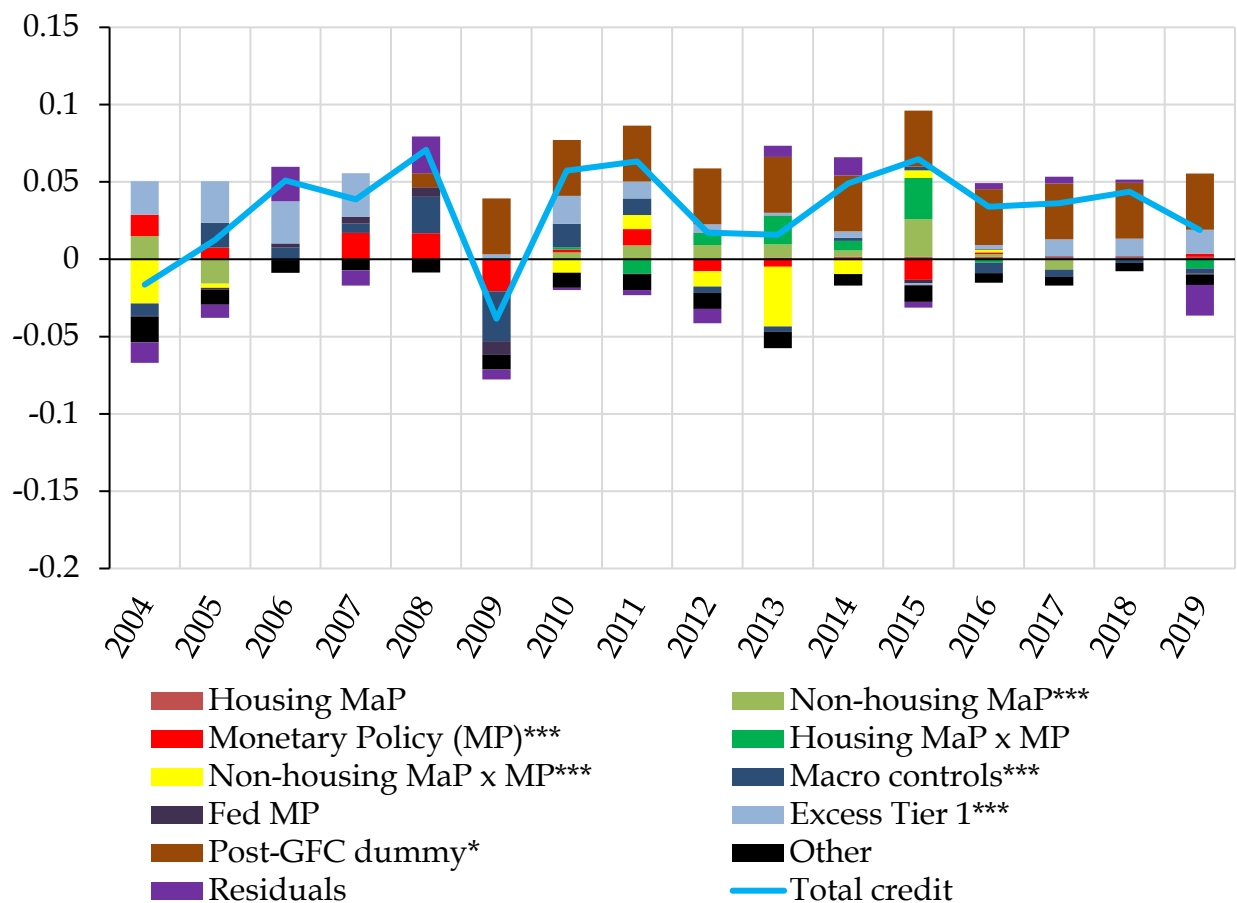

**Figure 11: Total Credit Decomposition**

*Notes*: Total credit is expressed in log difference (dlog). "Other" includes the intercept, bank fixed effects, seasonal dummies, data break dummies, and other financial ratios. ***, **, and * indicate significance at 1%, 5%, and 10% level, respectively. *Source*: Estimations, Table 6, Column 4.

Table 7 presents the main qualitative effect of MaP policy on each of the different credit components, based on the estimation including both the interaction terms and global monetary policy, presented in the fourth column of each table. We find that the effect of the various MaP measures on total credit is weaker than on some of its components due to the substitution effect mentioned earlier. Total bank credit is not affected significantly by the measures taken to curb housing credit or by the general MaP measures. The effect of concentration MaP measures is also weaker due to their influence in opposite directions on credit components.

We may conclude that the effect of MaP measures on total bank credit was less evident than on each of the credit components due to the substitution effects, probably originating from both the supply and demand sides. Housing MaP measures were aimed specifically at the housing market, and housing credit intended to minimize the risks to financial stability due to distress in this sector did succeed according to our estimation in moderating housing credit growth.

**Table 7: Summary of MaP Policy Qualitative Effect – Baseline Estimation**

|                          | Housing | Business | Consumer | Total       |
|--------------------------|---------|----------|----------|-------------|
| Housing MaP              | -***    | +**      | -        | +           |
| General MaP              | -       | -        | +        | +           |
| Concentration MaP        | +**     | -        | +        | +***/-***   |
| Housing MaP × MP         | +       | -        | -*       | -           |
| General MaP × MP         | +*      | +        | +        | +***        |
| Concentration MaP × MP   | +       | +        | +        | +**/-       |

*Notes*: ***, **, and * indicate significance at 1%, 5%, and 10% level, respectively. MP stands for monetary policy. Concentration MaP appears with 1 to 4 lags (+) and 4 to 7 lags (-) according to the specification for housing, business, and consumer credit. We refer to the estimation including the interaction term and the effect of US monetary policy, shown in Tables 4 to 6, Column 4.

Other MaP measures targeted at banks' capital structure may have achieved their goal of minimizing financial risks by inducing changes in the banks' balance sheets, while affecting credit growth only marginally.

## 5 Robustness Checks

We perform several robustness checks in order to verify our results.[15] First, we check whether aggregating general and concentration measures into a single index of dummies matters and find that qualitative results did not change. The substitution effect is still evident, as housing MaP measures tend to restrict housing credit growth but contribute to business credit growth and, on the whole, do not affect the evolution of total credit. The negative effect of housing MaP measures on consumer credit is significant in this alternative specification, representing possible complementarity between housing credit and consumer credit.

However, some of the effects of MaP and monetary policy surprises and their interactions were less significant for the housing and total credit estimations than for the business and consumer credit estimations. This result is expected since the effects in the benchmark specification of the general and concentration MaP measures are in opposite directions in the housing and total credit estimations. However, they are the same in the business and consumer credit estimations, although both insignificant. Table 8 presents the robustness checks aggregating both types of nonhousing MaP measures, including interactions and global monetary policy.

---

[15] We refer in this section to the results for the full specification (Column 4 of each table).

**Table 8: Robustness Checks, Sample 2004:Q2–2019:Q4, 7 Banks**

|  |  | 1 Housing credit | 2 Business credit | 3 Consumer credit | 4 Total credit |
|---|---|---|---|---|---|
| MaP | Cumulative MaP_housing(-1), since 2004 | -0.0013** | 0.001*** | -0.003* | 0.00 |
|  | MaP_non_housing (-1) to (-8) | -0.002 | -0.004** | 0.01 | 0.00 |
| Monetary Policy | Monetary policy(-1) until the GFC | -2.08 | 0.07 | 4.88 | 0.81 |
|  | Monetary policy(-2) until the GFC | -3.96*** | -3.22** | -20.25*** | -5.91*** |
|  | **Aggregated before 2008** | **-6.04*** | **-3.15** | **-15.37**** | **-5.11**** |
|  | Monetary policy(-1) after the GFC | -0.59 | 2.27 | 0.43 | 0.57 |
|  | Monetary policy(-2) after the GFC | 0.59 | 8.56*** | 2.48 | 3.71** |
|  | **Aggregated after 2008** | **0.01** | **10.83**** | **2.91** | **4.29*** |
| Interactions | Cumulative MaP_housing(-1) × Mon. policy(-1) | 0.24 | 0.03 | -1.43 | -0.04 |
|  | Cumulative MaP_housing(-1) × Mon. policy(-2) | -0.09 | -0.91** | -2.28* | -0.63** |
|  | **Aggregated MaP_housing × Mon. policy** | **0.15** | **-0.88** | **-3.71**** | **-0.68** |
|  | MaP_non_housing(-1) to (-8) × Mon. policy(-1) | 0.45 | 1.86 | 11.07* | 1.91 |
|  | MaP_non_housing (-1) to (-8) × Mon. policy(-2) | 1.75 | 0.45 | 12.03** | 2.24* |
|  | **Aggregated MaP_non_housing × Mon. policy** | **2.20** | **2.31** | **23.11**** | **4.16**** |
| Foreign | US monetary policy(-1) until the GFC | -1.06 | -1.26 | -10.00** | -2.28** |
|  | US monetary policy(-2) until the GFC | -2.09* | 1.25 | 7.06 | 1.18 |
|  | **US Aggregated before 2008** | **-3.15*** | **0.00** | **-2.94** | **-1.10** |
|  | US monetary policy(-1) after the GFC | 2.04 | -2.94 | -2.12 | -0.41 |
|  | US monetary policy(-2) after the GFC | -0.11 | -8.07** | -0.82 | -2.34 |
|  | **US Aggregated after 2008** | **1.93** | **-11.01**** | **-2.94** | **-2.74** |
| Banks^ | Excess Tier(-1) | 0.52*** |  | 0.14 | 0.31*** |
|  | Excess Tier(-5) |  | 0.36** |  |  |
| Macro | D(real wage(-1)) |  |  | 1.04*** |  |
|  | D(real wage(-2)) | 0.19** |  |  | -0.05 |
|  | D(unemployed 25-64(-2)) |  | -1.56*** | -0.97 | -1.11*** |
|  | D(unemployed 25-64(-5)) | -0.72** |  |  |  |
|  | D(BusinessGDP(-2)) |  | 0.17** |  |  |
|  | D(log(house_prices(-1)) | 0.09*** |  |  | -0.00 |
|  | Seasonal Dummies | Yes | Yes | Yes | Yes |
|  | Data Break Dummies | Yes | Yes | Yes | Yes |
|  | C | -0.001 | -0.02*** | 0.01 | -0.003 |
|  | CRISIS2008 | 0.014** | 0.01** | 0.01 | 0.01* |
|  | **Observations** | 441 | 441 | 441 | 441 |
|  | **Adjusted R²** | 0.17 | 0.31 | 0.07 | 0.28 |

*Notes*: ***, **, and * indicate significance at the 1%, 5%, and 10% level, respectively. ^ The estimation also includes the change from liabilities to assets, the change in the banks' real assets, the change in deposits as a share of assets, the change in liquid assets, and the change from excess reserves to assets, with varying lags. All results were insignificant.

In an additional robustness check, we omit from the estimation the insignificant MaP measures and find that our qualitative results do not change. We also experiment with competing models by including each of the MaP measures—housing, general, and concentration—as a cumulative index, compared to accumulating the general and concentration measures only over a fixed number of past periods in the benchmark specification. In this specification, the housing and business credit growth rates are not (significantly) affected by MaP measures or their interaction with monetary policy, except for the interaction of monetary policy with the housing MaP measures in the business credit sector, which becomes significantly negative.

According to this specification, the interaction between accommodative monetary policy and housing MaP measures tends to increase business credit. In other words, supply-side substitution effects lead banks to increase the supply of business credit more when housing market MaP measures are accompanied by accommodative monetary policy. For consumer credit, we find that housing MaP measures become significantly positive and increase consumer growth. By contrast, general MaP measures accumulated from the beginning of the estimated period become significantly negative and tend to reduce it. Interactions do not significantly affect consumer credit growth in this specification.

The last robustness check includes the relative surprise to the interest rate (surprise divided by the level of the interest rate) in order to take into account the absolute level of the interest rate relative to the size of the surprise. The results remain qualitatively unchanged,[16] although the interaction terms with MaP policy become insignificant.

# 6  Concluding Remarks

This paper analyzes the effect of monetary and MaP policies on bank credit. We present the results of an analysis of the effect of domestic monetary policy and MaP policy implemented in Israel during the period 2004–19 on domestic bank credit. We estimate the effect of monetary policy, MaP measures, and the interaction between them on housing credit, consumer credit, credit to the business sector, and total credit.

We find that the MaP policy affected the development of bank credit. Our results show that MaP measures targeted at housing credit did reduce the growth rate of this credit while contributing to the increased growth of business credit. General MaP measures worked to reduce the growth rate of business credit. We could not find a clear effect of the single measure targeted at the lending concentration on any of the credit components.

Monetary policy was found to have a clear and significant effect on housing and total credit growth rates before the GFC, but this effect did not persist after the crisis, at least according to our estimation. We find some evidence for the effect of monetary policy when it interacts with MaP measures that were taken after 2008. As the share of foreign currency credit is very small, the effect of monetary policy in the US is in the same direction as that of domestic monetary policy, indicating that it may be a signal for possible future local policy rather than a substitute.

We find that the interaction between monetary policy and MaP policy is important. We show that accommodative monetary policy tended to increase consumer credit and total credit in some specifications, when interacting with macroprudential policies targeting the housing market.

We also find that macroeconomic conditions, i.e., output growth and employment growth, are associated with credit growth rates, and that bank-specific characteristics, particularly the bank's capital relative to the supervisory requirements, affect the supply of credit.

---

[16] In this specification, there is no evidence for the substitution effect of housing MaP measures between housing credit growth and business credit growth.

# Appendix – Estimation Including ECB Monetary Policy Surprises

**Table 3a: Housing Credit, Sample 2004:Q2 – 2019:Q4, 7 Banks, Incl. ECB Monetary Policy**

| | | 1 Basic – excl. interactions and global policy | 2 Including interactions | 3 Including global policy | 4 Including interactions and global monetary policy |
|---|---|---|---|---|---|
| MaP | Cumulative MaP_housing(-1), since 2004 | -0.001** | -0.002** | -0.002*** | -0.002*** |
| | MaP_general(-1) to (-8) | -0.004* | -0.002 | -0.004* | -0.003 |
| | MaP_concentration(-1) to (-4) | 0.002 | 0.006 | 0.005 | 0.01** |
| Monetary Policy | Monetary policy(-1) until the GFC | -1.89* | -2.23** | -0.81 | -0.86 |
| | Monetary policy(-2) until the GFC | -4.56*** | -4.53*** | -3.13** | -3.61*** |
| | **Aggregated before 2008** | **-6.45***** | **-6.76***** | **-3.94**** | **-4.46**** |
| | Monetary policy(-1) after the GFC | 0.82 | 0.88 | -0.35 | -3.31 |
| | Monetary policy(-2) after the GFC | 0.75 | 0.56 | 0.31 | 0.64 |
| | **Aggregated after 2008** | **1.57*** | **1.43** | **0.05** | **-2.67** |
| Interactions | Cumulative MaP_housing(-1) × Mon. policy(-1) | | -0.14 | | 0.29 |
| | Cumulative MaP_housing(-1) × Mon. policy(-2) | | -0.22 | | -0.24 |
| | **Aggregated MaP_housing × Mon. policy** | | **-0.36** | | **0.05** |
| | MaP_general(-1) to (-8) × Mon. policy(-1) | | 1.04 | | 3.22* |
| | MaP_general(-1) to (-8) × Mon. policy(-2) | | 2.76 | | 2.43 |
| | **Aggregated MaP_general × Mon. policy** | | **3.80** | | **5.65*** |
| | MaP_concentr.(-1) to (-4) × Mon. policy(-1) | | 4.08 | | 2.59 |
| | MaP_concentr.(-1) to (-4) × Mon. policy(-2) | | 3.09 | | 5.07* |
| | **Aggregated MaP_concentr. × Mon. policy** | | **7.18*** | | **7.66*** |
| Foreign | US monetary policy(-1) until the GFC | | | -2.58** | -2.53* |
| | US monetary policy(-2) until the GFC | | | -2.44* | -2.08 |
| | **US Aggregated before 2008** | | | **-5.02***** | **-4.61**** |
| | US monetary policy(-1) after the GFC | | | -0.27 | 4.97 |
| | US monetary policy(-2) after the GFC | | | 0.69 | 0.96 |
| | **US Aggregated after 2008** | | | **0.41** | **5.93** |
| | ECB monetary policy(-1) until the GFC | | | 24.93** | 32.51** |
| | ECB monetary policy(-2) until the GFC | | | -15.00 | -15.14 |
| | **ECB Aggregated before 2008** | | | **9.93** | **17.37** |
| | ECB monetary policy(-1) after the GFC | | | 2.96 | 2.22 |
| | ECB monetary policy(-2) after the GFC | | | -1.02 | --0.94 |
| | **ECB Aggregated after 2008** | | | **1.94** | **1.27** |
| Banks^ | Excess Tier(-1) | 0.38*** | 0.40*** | 0.59*** | 0.61*** |
| Macro | D(real wage(-2)) | 0.12* | 0.14** | 0.18** | 0.16* |
| | D(unemployed 25-64(-5)) | -0.77*** | -0.61** | -0.64** | -0.57* |
| | D(log(house_prices(-1))) | 0.07*** | 0.06** | 0.05** | 0.05* |
| | Seasonal Dummies | Yes | Yes | Yes | Yes |
| | Data Break Dummies | Yes | Yes | Yes | Yes |
| | C | 0.001 | 0.001 | -0.003 | -0.001 |
| | CRISIS2008 | 0.01** | 0.02*** | 0.02*** | 0.02*** |
| | **Observations** | 441 | 441 | 441 | 441 |
| | **Adjusted R²** | 0.17 | 0.17 | 0.19 | 0.20 |

*Notes*: *p-value < 10%, **p-value < 5%, ***p-value < 1%. ^ The estimation also includes the change from liabilities to assets, the change in the banks' real assets, the change in deposits as a share of assets, the change in liquid assets, and the change from excess reserves to assets, with varying lags. All results were insignificant.

# Table 4a: Business Credit, Sample 2004:Q2 – 2019:Q4, 7 Banks, Incl. ECB Monetary Policy

|  |  | 1<br>Basic – excl. interactions and global policy | 2<br>Including interactions | 3<br>Including global policy | 4<br>Including interactions and global monetary policy |
|---|---|---|---|---|---|
| MaP | Cumulative MaP_housing(-1), since 2004 | 0.001** | 0.001** | 0.001** | 0.001** |
|  | MaP_general(-1) to (-8) | -0.008*** | -0.005* | -0.006** | -0.005 |
|  | MaP_concentration(-4) to (-7) | -0.004 | -0.006 | -0.004 | -0.004 |
| Monetary Policy | Monetary policy(-1) until the GFC | 0.12 | -0.44 | 0.45 | -0.81 |
|  | Monetary policy(-2) until the GFC | -2.75** | -2.17* | -3.93*** | -2.69 |
|  | **Aggregated before 2008** | **-2.63*** | **-2.61** | **-3.48*** | **-3.50*** |
|  | Monetary policy(-1) after the GFC | 1.28** | 0.22 | 3.00*** | 2.68 |
|  | Monetary policy(-2) after the GFC | 3.32*** | 3.80*** | 3.45*** | 6.33** |
|  | **Aggregated after 2008** | **4.60*** | **4.02*** | **6.44*** | **9.01**** |
| Interactions | Cumulative MaP_housing(-1) × Mon. policy(-1) |  | 0.23 |  | -0.02 |
|  | Cumulative MaP_housing(-1) × Mon. policy(-2) |  | -0.43* |  | -0.80** |
|  | **Aggregated MaP_housing × Mon. policy** |  | **-0.20** |  | **-0.83** |
|  | MaP_general(-1) to (-8) × Mon. policy(-1) |  | 3.00 |  | 1.75 |
|  | MaP_general(-1) to (-8) × Mon. policy(-2) |  | 3.57* |  | 3.26 |
|  | **Aggregated MaP_general × Mon. policy** |  | **6.57*** |  | **5.01** |
|  | MaP_concentr.(-4) to (-7) × Mon. policy(-1) |  | 1.89 |  | 1.03 |
|  | MaP_concentr.(-4) to (-7) × Mon. policy(-2) |  | -2.66 |  | -2.85 |
|  | **Aggregated MaP_concentr. × Mon. policy** |  | **-0.77** |  | **-1.81** |
| Foreign | US monetary policy(-1) until the GFC |  |  | -1.40 | -0.50 |
|  | US monetary policy(-2) until the GFC |  |  | 1.72 | 0.84 |
|  | **US Aggregated before 2008** |  |  | **0.33** | **0.34** |
|  | US monetary policy(-1) after the GFC |  |  | -2.40 | -1.01 |
|  | US monetary policy(-2) after the GFC |  |  | 0.05 | -4.68 |
|  | **US Aggregated after 2008** |  |  | **-2.35** | **-5.68** |
|  | ECB monetary policy(-1) until the GFC |  |  | -21.19 | -23.98 |
|  | ECB monetary policy(-2) until the GFC |  |  | -13.12 | -15.72 |
|  | **ECB Aggregated before 2008** |  |  | **-34.31*** | **-39.70*** |
|  | ECB monetary policy(-1) after the GFC |  |  | -2.80 | -3.69 |
|  | ECB monetary policy(-2) after the GFC |  |  | 0.02 | 0.80 |
|  | **ECB Aggregated after 2008** |  |  | **-2.78** | **-2.90** |
| Banks^ | Excess Tier(-5) | 0.40*** | 0.29* | 0.33** | 0.39** |
|  | D(Liabilities(-1)/Assets(-1)) | -0.73 | -1.06* | -1.02* | -1.21** |
| Macro | D(BusinessGDP(-2)) | 0.22*** | 0.20*** | 0.17* | 0.17* |
|  | D(unemployed 25-64(-2)) | -1.55*** | -1.77*** | -1.79*** | -1.66*** |
|  | Seasonal Dummies | Yes | Yes | Yes | Yes |
|  | Data Break Dummies | Yes | Yes | Yes | Yes |
|  | C | -0.03** | -0.02** | -0.03*** | -0.02*** |
|  | CRISIS2008 | 0.02*** | 0.01*** | 0.01*** | 0.02*** |
|  | **Observations** | 441 | 441 | 441 | 441 |
|  | **Adjusted R²** | 0.30 | 0.30 | 0.30 | 0.30 |

*Notes*: *p-value < 10%, **p-value < 5%, ***p-value < 1%. ^ The estimation also includes the change from liabilities to assets, the change in the banks' real assets, the change in deposits as a share of assets, the change in liquid assets, and the change from excess reserves to assets, with varying lags. All results were insignificant.

**Table 5a: Consumer Credit, Sample 2004:Q2 – 2019:Q4, 7 Banks, Incl. ECB Monetary Policy**

|  |  | 1<br>Basic – excl. interactions and global policy | 2<br>Including interactions | 3<br>Including global policy | 4<br>Including interactions and global monetary policy |
|---|---|---|---|---|---|
| MaP | Cumulative MaP_housing(-1), since 2004 | -0.001 | -0.002 | -0.002 | -0.003 |
|  | MaP_general(-1) to (-8) | 0.003 | 0.005 | 0.001 | 0.005 |
|  | MaP_concentration(-1) to (-4) | -0.003 | 0.013 | 0.002 | 0.01 |
| Monetary Policy | Monetary policy(-1) until the GFC | -0.52 | -1.60 | 7.32 | 6.95 |
|  | Monetary policy(-2) until the GFC | -12.37*** | -12.73*** | -20.03*** | -20.77*** |
|  | **Aggregated before 2008** | **-12.89**** | **-14.33**** | **-12.71*** | **-13.82**** |
|  | Monetary policy(-1) after the GFC | -0.07 | -0.77 | 1.34 | 1.29 |
|  | Monetary policy(-2) after the GFC | 0.54 | 3.07 | -1.46 | 12.52 |
|  | **Aggregated after 2008** | **0.47** | **2.30** | **0.12** | **13. 81** |
| Interactions | Cumulative MaP_housing(-1) × Mon. policy(-1) |  | -0.80 |  | -0.87 |
|  | Cumulative MaP_housing(-1) × Mon. policy(-2) |  | -2.05** |  | -2.91** |
|  | **Aggregated MaP_housing × Mon. policy** |  | **-2.84**** |  | **-3.79**** |
|  | MaP_general(-1) to (-8) × Mon. policy(-1) |  | 6.2 |  | 7.11 |
|  | MaP_general(-1) to (-8) × Mon. policy(-2) |  | 7.21 |  | 2.72 |
|  | **Aggregated MaP_general × Mon. policy** |  | **13.41** |  | **9.83** |
|  | MaP_concentr.(-1) to (-4) × Mon. policy(-1) |  | 7.88 |  | 0.22 |
|  | MaP_concentr.(-1) to (-4) × Mon. policy(-2) |  | 19.06 |  | 21.73* |
|  | **Aggregated MaP_concentr. × Mon. policy** |  | **26.94*** |  | **21.95** |
| Foreign | US monetary policy(-1) until the GFC |  |  | -10.65** | -11.07** |
|  | US monetary policy(-2) until the GFC |  |  | 9.45** | 10.05** |
|  | **US Aggregated before 2008** |  |  | **-1.20** | **-1.02** |
|  | US monetary policy(-1) after the GFC |  |  | 2.65 | 3.55 |
|  | US monetary policy(-2) after the GFC |  |  | 14.05* | -6.68 |
|  | **US Aggregated after 2008** |  |  | **16.69** | **-3.13** |
|  | ECB monetary policy(-1) until the GFC |  |  | 9.39 | 52.96 |
|  | ECB monetary policy(-2) until the GFC |  |  | -114.43** | -114.87** |
|  | **ECB Aggregated before 2008** |  |  | **-105.04** | **-61.91** |
|  | ECB monetary policy(-1) after the GFC |  |  | -7.44 | -6.63 |
|  | ECB monetary policy(-2) after the GFC |  |  | -9.76 | -9.58 |
|  | **ECB Aggregated after 2008** |  |  | **-17.19** | **-16.22** |
| Banks^ | Excess Tier(-1) | -0.15 | 0.05 | -0.38 | -0.24 |
|  | D(Liabilities(-1)/Assets(-1)) | -3.96** | -4.29** | -5.29*** | -5.09*** |
| Macro | D(real wage(-1)) | 0.66*** | 0.63** | 1.13*** | 0.99*** |
|  | D(unemployed 25-64(-2)) | -0.93 | -0.31 | -1.06 | 0.19 |
|  | Seasonal Dummies | Yes | Yes | Yes | Yes |
|  | Data Break Dummies | Yes | Yes | Yes | Yes |
|  | C | 0.03** | 0.03* | 0.020 | 0.03* |
|  | CRISIS2008 | -0.01 | -0.003 | 0.01 | 0.000 |
|  | **Observations** | 441 | 441 | 441 | 441 |
|  | **Adjusted R²** | 0.04 | 0.04 | 0.06 | 0.06 |

*Notes*: *p-value < 10%, **p-value < 5%, ***p-value < 1%. ^ The estimation also includes the change from liabilities to assets, the change in the banks' real assets, the change in deposits as a share of assets, the change in liquid assets, and the change from excess reserves to assets, with varying lags. All results were insignificant.

**Table 6a: Total Credit, Sample 2004:Q2 – 2019:Q4, 7 Banks, Incl. ECB Monetary Policy**

| | | 1<br>Basic – excl. interactions and global policy | 2<br>Including interactions | 3<br>Including global policy | 4<br>Including interactions and global monetary policy |
|---|---|---|---|---|---|
| MaP | Cumulative MaP_housing(-1), since 2004 | 0.0002 | -0.0002 | 0.0005 | -0.0001 |
| | MaP_general(-1) to (-8) | -0.0002 | 0.002 | 0.0004 | 0.002 |
| | MaP_concentration(-1) to (-4) | 0.006** | 0.008** | 0.008** | 0.009** |
| | MaP_concentration(-4) to (-7) | -0.008*** | -0.009*** | -0.007*** | -0.007*** |
| Monetary Policy | Monetary policy(-1) until the GFC | -0.10 | -0.68 | 1.66** | 1.17 |
| | Monetary policy(-2) until the GFC | -4.60*** | -4.74*** | -6.20*** | -6.53*** |
| | **Aggregated before 2008** | **-4.70*** | **-5.42*** | **-4.54*** | **-5.36*** |
| | Monetary policy(-1) after the GFC | 0.59 | 0.17 | 1.23 | 0.13 |
| | Monetary policy(-2) after the GFC | 1.76*** | 2.31*** | 1.40* | 3.19* |
| | **Aggregated after 2008** | **2.35*** | **2.48*** | **2.62*** | **3.32** |
| Interactions | Cumulative MaP_housing(-1) × Mon. policy(-1) | | -0.12 | | -0.08 |
| | Cumulative MaP_housing(-1) × Mon. policy(-2) | | -0.62*** | | -0.72** |
| | **Aggregated MaP_housing × Mon. policy** | | **-0.74*** | | **-0.79*** |
| | MaP_general(-1) to (-8) × Mon. policy(-1) | | 2.69** | | 2.82* |
| | MaP_general(-1) to (-8) × Mon. policy(-2) | | 4.31*** | | 4.01*** |
| | **Aggregated MaP_general × Mon. policy** | | **6.99*** | | **6.83*** |
| | MaP_concentr.(-1) to (-4) × Mon. policy(-1) | | 7.83*** | | 6.25*** |
| | MaP_concentr.(-1) to (-4) × Mon. policy(-2) | | -0.17 | | 0.03 |
| | **Aggregated MaP_concentr. × Mon. policy** | | **7.66*** | | **6.29*** |
| | MaP_concentr.(-4) to (-7) × Mon. policy(-1) | | -2.04 | | -3.14 |
| | MaP_concentr.(-4) to (-7) × Mon. policy(-2) | | 1.49 | | 3.34 |
| | **Aggregated MaP_concentr. × Mon. policy** | | **-0.55** | | **0.20** |
| Foreign | US monetary policy(-1) until the GFC | | | -2.44** | -1.96* |
| | US monetary policy(-2) until the GFC | | | 2.04* | 2.03* |
| | **US Aggregated before 2008** | | | **-0.41** | **0.07** |
| | US monetary policy(-1) after the GFC | | | -0.81 | 1.59 |
| | US monetary policy(-2) after the GFC | | | 2.34 | -0.72 |
| | **US Aggregated after 2008** | | | **1.52** | **0.87** |
| | ECB monetary policy(-1) until the GFC | | | -7.80 | 1.13 |
| | ECB monetary policy(-2) until the GFC | | | -21.86** | -28.59*** |
| | **ECB Aggregated before 2008** | | | **-29.66*** | **-2746*** |
| | ECB monetary policy(-1) after the GFC | | | -0.93 | -1.75 |
| | ECB monetary policy(-2) after the GFC | | | -1.58 | -0.21 |
| | **ECB Aggregated after 2008** | | | **-2.50** | **-1.96** |
| Banks^ | Excess Tier(-1) | 0.27*** | 0.32*** | 0.26** | 0.29** |
| | D(log(Real Assets(-1))) | 0.08 | 0.08* | 0.08 | 0.09* |
| Macro | YoY(Real Wage(-2)) | 0.01 | 0.04 | -0.08 | -0.02 |
| | D(Unemployed 25-64(-2)) | -1.15*** | -1.01*** | -1.31*** | -0.95*** |
| | YoY(House Prices(-1)) | 0.01 | 0.001 | 0.007 | -0.007 |
| | Seasonal Dummies | Yes | Yes | Yes | Yes |
| | Data Break Dummies | Yes | Yes | Yes | Yes |
| | C | -0.002 | -0.002 | -0.003 | -0.001 |
| | CRISIS2008 | 0.007 | 0.009* | 0.007 | 0.01* |
| | **Observations** | 441 | 441 | 441 | 441 |
| | **Adjusted R²** | 0.28 | 0.32 | 0.29 | 0.32 |

*Notes*: *p-value < 10%, **p-value < 5%, ***p-value < 1%. ^ The estimation also includes the change from liabilities to assets, the change in the banks' real assets, the change in deposits as a share of assets, the change in liquid assets, and the change from excess reserves to assets, with varying lags. All results were insignificant.